\documentclass[12pt,preprint,natbib]{aastex}
\usepackage[utf8]{inputenc}
\usepackage{caption}
\usepackage{natbib}

\newcommand{\gpiwave}{$0.957-1.120$} 
\newcommand{\gpiysattostar}{$2.345\times10^{-4}$} 
\newcommand{\distance}{$103.1\pm5.3$}

\newcommand{\gpishortcutoff}{$0.953$} 
\newcommand{\gpilongcutoff}{$1.12$} 

\newcommand{\scexaoywave}{$0.97-1.07$} 
\newcommand{\scexaojwave}{$1.17-1.33$} 
\newcommand{\scexaohwave}{$1.49-1.78$} 

\newcommand{\absmagy}{$10.23\pm0.11$}
\newcommand{\absmagj}{$9.63\pm0.10$}
\newcommand{\absmagh}{$9.14\pm0.10$}
\newcommand{\absjmagc}{$8.29\pm0.10$}
\newcommand{\abshmagc}{$7.75\pm0.10$}
\newcommand{\absymagc}{$8.67\pm0.10$}

\newcommand{\yjcolor}{$0.61\pm0.06$}

\newcommand{\jhcolor}{$0.48\pm0.05$}

\newcommand{\scexaoyb}{$15.30\pm0.04$}
\newcommand{\scexaojb}{$14.69\pm0.05$}
\newcommand{\scexaohb}{$14.21\pm0.02$}

\newcommand{\scexaoyc}{$13.74\pm0.02$}
\newcommand{\scexaojc}{$13.37\pm0.02$}
\newcommand{\scexaohc}{$12.81\pm0.02$}

\newcommand{\nicijb}{$15.83\pm0.10$}
\newcommand{\nicihb}{$14.65\pm0.08$}
\newcommand{\nicikb}{$14.12\pm0.05$}

\newcommand{\nicijc}{$13.31\pm0.04$}
\newcommand{\nicihc}{$12.54\pm0.03$}
\newcommand{\nicikc}{$12.18\pm0.06$}

\newcommand{\gpiyb}{$15.13\pm0.21$} 
\newcommand{\vltyb}{$15.19\pm0.02$} 
\newcommand{\vltjb}{$14.47\pm0.11$} 
\newcommand{\vlthb}{$13.97\pm0.08$} 
\newcommand{\vltkb}{$13.96\pm0.12$}

\newcommand{\keckmb}{$13.81\pm0.24$}

\newcommand{\deltamagy}{$8.34\pm0.04$}
\newcommand{\deltamagj}{$7.72\pm0.02$}
\newcommand{\deltamagh}{$7.20\pm0.02$}

\newcommand{\deltamagyc}{$6.78\pm0.01$}
\newcommand{\deltamagjc}{$6.38\pm0.01$}
\newcommand{\deltamaghc}{$5.80\pm0.01$}

\newcommand{\bolocorjc}{$1.875\pm0.035$}
\newcommand{\boloc}{$10.16\pm0.11$}%

\newcommand{\normphot}{$0.98-1.42$}
\newcommand{\normspec}{$1.22-1.37$}

\newcommand{\teffa}{$9011\pm85$}
\newcommand{\teffb}{$3000-3100$} 
\newcommand{\teffc}{$3260\pm100$}
\newcommand{\spta}{A$0$}
\newcommand{\sptb}{M$5.5^{+1.0}_{-0.5}$} 
\newcommand{\sptc}{M$3.5\pm0.5$} 
\newcommand{\loglsuna}{$1.12\pm0.07$}
\newcommand{\loglsunb}{$-2.76\pm0.05$}
\newcommand{\loglsunc}{$-2.16\pm0.04$}
\newcommand{\logga}{} 
\newcommand{\loggb}{$4.0-4.5$} 
\newcommand{\loggc}{}

\newcommand{\massb}{70--90} 
\newcommand{\massc}{175--200}

\newcommand{\bestfitradius}{$1.55\pm0.10$} 

\newcommand{\chisqrbestphot}{$0.63$}
\newcommand{\teffbestphot}{$2900-3500$}
\newcommand{\loggbestphot}{$3.5-5.5$}

\newcommand{\chisqrbestspec}{$0.92$}

\newcommand{\ndy}{$0.388\pm0.008$\%}
\newcommand{\ndj}{$0.590\pm0.004$\%}
\newcommand{\ndh}{$0.854\pm0.002$\%}

\title{ SCExAO and GPI $YJH$ Band Photometry and Integral Field Spectroscopy of the Young Brown Dwarf Companion to HD 1160}
\author{
E. Victor Garcia\altaffilmark{1,2,36}, 
Thayne Currie\altaffilmark{3}, 
Olivier Guyon\altaffilmark{3,4,5}, 
Keivan G.\ Stassun\altaffilmark{2,6}, 
Nemanja Jovanovic\altaffilmark{3}, 
Julien Lozi\altaffilmark{3}, 
Tomoyuki Kudo\altaffilmark{3}, 
Danielle Doughty\altaffilmark{7},
Josh Schlieder\altaffilmark{8},
J.~Kwon\altaffilmark{9},
T.~Uyama\altaffilmark{10},
M.~Kuzuhara\altaffilmark{4,11,12},  
J.~C.~Carson\altaffilmark{13}, 
T.~Nakagawa\altaffilmark{9},
J.~Hashimoto\altaffilmark{4,11},
N.~Kusakabe\altaffilmark{4,11},
L.~Abe\altaffilmark{14}, 
W.~Brandner\altaffilmark{15}, 
T.~D.~Brandt\altaffilmark{16,17},
M.~Feldt\altaffilmark{15},
M.~Goto\altaffilmark{18},
C.~A.~Grady\altaffilmark{19,20,21},
Y.~Hayano\altaffilmark{3},
M.~Hayashi\altaffilmark{11},
S.~S.~Hayashi\altaffilmark{3,22},
T.~Henning\altaffilmark{14},
K.~W.~Hodapp\altaffilmark{23},
M.~Ishii\altaffilmark{11},
M.~Iye\altaffilmark{11},
M.~Janson\altaffilmark{24},
R.~Kandori\altaffilmark{11},
G.~R.~Knapp\altaffilmark{25},
T.~Matsuo\altaffilmark{26},
M.~W.~McElwain\altaffilmark{18},
S.~Miyama\altaffilmark{27},
J.-I.~Morino\altaffilmark{11},
A.~Moro-Martin\altaffilmark{28,29},
T.~Nishimura\altaffilmark{3},
T.-S.~Pyo\altaffilmark{3},
E.~Serabyn\altaffilmark{30},
T.~Suenaga\altaffilmark{10,22},
H.~Suto\altaffilmark{4,11},
R.~Suzuki\altaffilmark{11},
Y.~H.~Takahashi\altaffilmark{11,30},
H.~Takami\altaffilmark{31},
M.~Takami\altaffilmark{31},
N.~Takato\altaffilmark{3},
H.~Terada\altaffilmark{11},
C.~Thalmann\altaffilmark{32},
E.~L.~Turner\altaffilmark{25,33},
M.~Watanabe\altaffilmark{34},
J.~Wisniewski\altaffilmark{35},
T.~Yamada\altaffilmark{9},
T.~Usuda\altaffilmark{11},
and M.~Tamura\altaffilmark{4,10,11}}

\altaffiltext{1}{Lowell Observatory, Flagstaff, AZ, 86001, USA; eugenio.v.garcia@gmail.com}
\altaffiltext{2}{Department of Physics and Astronomy, Vanderbilt University, Nashville, TN}
\altaffiltext{36}{Lawrence Livermore National Laboratory, L-210, 7000 East Avenue, Livermore, CA 94550, USA}

\altaffiltext{3}{Subaru Telescope, National Astronomical Observatory of Japan, 650 N.~A'ohoku Place, Hilo, HI 96720, USA}
\altaffiltext{4}{Astrobiology Center of NINS, 2-21-1 Osawa, Mitaka, Tokyo, 181-8588, Japan}
\altaffiltext{5}{Steward Observatory, University of Arizona, 933 N Cherry Ave., Tucson, AZ 85721, USA}
\altaffiltext{6}{Department of Physics, Fisk University, Nashville, TN 37208, USA}
\altaffiltext{7}{College of Optical Sciences, University of Arizona, Tucson, AZ 85721}
\altaffiltext{8}{NASA Postdoctoral Program Fellow, NASA Ames Research Center, Moffett Field, CA}
\altaffiltext{9}{Department of Space Astronomy and Astrophysics, Institute of Space \& Astronautical Science (ISAS), Japan Aerospace Exploration Agency (JAXA), 3-1-1 Yoshinodai, Chuo-ku, Sagamihara, Kanagawa 252-5210, Japan}
\altaffiltext{10}{Department of Astronomy, University of Tokyo, 7-3-1, Hongo, Bunkyo-ku, Tokyo, 113-0033, Japan}
\altaffiltext{11}{National Astronomical Observatory of Japan, 2-21-1, Osawa, Mitaka, Tokyo, 181-8588, Japan}
\altaffiltext{12}{Department of Earth and Planetary Sciences, Tokyo Institute of 
Technology, Ookayama, Meguro-ku, Tokyo, 152-8551, Japan}
\altaffiltext{13}{Department of Physics and Astronomy, College of Charleston, 58 Coming St., Charleston, SC 29424, USA}
\altaffiltext{14}{Laboratoire Lagrange (UMR 7293), Universit\'e de Nice-Sophia Antipolis, CNRS, Observatoire de la C\^{o}te d'Azur, 28 Avenue Valrose, F-06108 Nice Cedex 2, France}
\altaffiltext{15}{Max Planck Institute for Astronomy, K\"{o}nigstuhl 17, 69117 Heidelberg, Germany}
\altaffiltext{16}{Astrophysics Department, Institute for Advanced Study, Princeton, NJ, USA}
\altaffiltext{17}{NASA Sagan Fellow}
\altaffiltext{18}{Universit\"{a}ts-Sternwarte M\"{u}nchen, Ludwig-Maximilians-Universit\"{a}t, Scheinerstr. 1, 81679 M\"{u}nchen,Germany}
\altaffiltext{19}{Exoplanets and Stellar Astrophysics Laboratory, Code 667, Goddard Space Flight Center, Greenbelt, MD 20771, USA}
\altaffiltext{20}{Eureka Scientific, 2452 Delmer, Suite 100, Oakland, CA 96002, USA}
\altaffiltext{21}{Goddard Center for Astrobiology}
\altaffiltext{22}{Department of Astronomical Science, SOKENDAI (The Graduate University for Advanced Studies), 2-21-1 Osawa, Mitaka, Tokyo, 181-8588, Japan}
\altaffiltext{23}{Institute for Astronomy, University of Hawaii, 640 N.~A'ohoku Place, Hilo, HI 96720, USA}
\altaffiltext{24}{Department of Astronomy, Stockholm University, AlbaNova University Center, SE-10691, Stockholm, Sweden}
\altaffiltext{25}{Department of Astrophysical Science, Princeton University, Peyton Hall, Ivy Lane, Princeton, NJ 08544, USA}
\altaffiltext{26}{Department of Earth and Space Science, Graduate School of Science, Osaka University, 1-1 Machikaneyamacho, Toyonaka, Osaka, 560-0043, Japan}
\altaffiltext{27}{Hiroshima University, 1-3-2, Kagamiyama, Higashihiroshima, Hiroshima, 739-8511, Japan}
\altaffiltext{28}{Space  Telescope  Science  Institute,  3700  San  Martin  Drive,  Baltimore, MD 21218, USA} 
\altaffiltext{29}{Center for Astrophysical Sciences, Johns Hopkins University, Baltimore, MD 21218, USA} 
\altaffiltext{30}{Jet Propulsion Laboratory, California Institute of Technology, Pasadena, CA, 171-113, USA}

\altaffiltext{31}{Institute of Astronomy and Astrophysics, Academia Sinica, P.O. Box 23-141, Taipei 10617, Taiwan}
\altaffiltext{32}{Swiss Federal Institute of Technology (ETH Zurich), Institute for Astronomy,Wolfgang-Pauli-Strasse 27, CH-8093 Zurich, Switzerland}
\altaffiltext{33}{Kavli Institute for Physics and Mathematics of the Universe, The University of Tokyo, 5-1-5, Kashiwanoha, Kashiwa, Chiba, 277-8568, Japan}
\altaffiltext{34}{Department of Cosmosciences, Hokkaido University, Kita-ku, Sapporo, Hokkaido, 060-0810, Japan}
\altaffiltext{35}{H.~L.~Dodge Department of Physics \& Astronomy, University of Oklahoma, 440 W Brooks St Norman, OK 73019, USA}

\begin{document}

\begin{abstract}
We present high signal-to-noise ratio, precise $YJH$ photometry and $Y$ band (\gpiwave~$\mu$m) spectroscopy of HD 1160 B, a young substellar companion discovered from the Gemini NICI Planet Finding Campaign, using the Subaru Coronagraphic Extreme Adaptive Optics instrument and the Gemini Planet Imager.   HD 1160 B has typical mid-M dwarf-like infrared colors and a spectral type of M5.5$^{+1.0}_{-0.5}$, where the blue edge of our $Y$ band spectrum rules out earlier spectral types.  Atmospheric modeling suggests HD 1160 B having an effective temperature of 3000--3100 $K$, a surface gravity of log $g$ = 4--4.5, a radius of~\bestfitradius~$R_{\rm J}$, and a luminosity of log $L$/$L_{\odot} = -2.76 \pm 0.05$.   Neither the primary's Hertzspring-Russell diagram position nor atmospheric modeling of HD 1160 B show evidence for a sub-solar metallicity.   The interpretation of the HD 1160 B depends on which stellar system components are used to estimate an age.  Considering HD 1160 A, B and C jointly, we derive an age of 80--125 Myr, implying that HD 1160 B straddles the hydrogen-burning limit (70--90 $M_{\rm J}$).  If we consider HD 1160 A alone, younger ages (20--125 Myr) and a brown dwarf-like mass (35--90 $M_{\rm J}$) are possible.  Interferometric measurements of the primary, a precise GAIA parallax, and moderate resolution spectroscopy can better constrain the system's age and how HD 1160 B fits within the context of (sub)stellar evolution.

\end{abstract}

\section{Introduction}
In addition to about a dozen superjovian-mass extrasolar planets \citep[e.g.][]{Marois2008,Lagrange2010,KrausIreland2012,Rameau2013,Quanz2013,Kuzuhara2013,Carson2013,Currie14a,Currie15a, Macintosh2015}, the past decade of ground-based \textit{direct imaging} observations of young stars have also revealed numerous higher mass (ratio) and (often) wider separation brown dwarf companions \citep{Chauvin2005,Biller2010,Ireland2011}.  Their typically more modest companion-to-primary contrasts make their detection and precise photometric/spectroscopic calibration possible without having to correct for throughput/signal losses inherent in the algorithms needed to detect more extreme contrast-ratio exoplanets \citep[e.g.][]{Marois2010spie}.  
They, and free-floating brown dwarfs of comparable ages, can reveal general trends in infrared colors and spectra that clarify the properties of young substellar atmospheres, can serve as important tests of the same atmosphere models used to characterize bona fide imaged exoplanets, and may provide insights into the formation of substellar objects \citep[e.g.][]{Faherty16,Kratter2010,Reggiani2011}.  

One particularly interesting system with a brown dwarf companion is HD 1160, located at a distance of~\distance~pc \citep{vanLeeuwen07} with an A0V primary that is comparable in mass to at least some stars around which planets have been imaged (e.g. $\beta$ Pic).  Targeted by the NICI Planet-Finding Campaign \citep{Liu10}, HD 1160 is orbited by a pair of substellar companions: HD 1160 B at $\sim$ 80 au and HD 1160 C at 530 au \citep[][hereafter N12]{Nielsen12}.  

N12's photometry for HD 1160 B and spectrum for HD 1160 C suggested that the former is a $L0\pm2$ brown dwarf and the latter a M3.5 star.  N12 used HD 1160 B's and C's red $J-K_s$ colors, HD 1160 A's luminosity, and the system's large space motion to argue that all components are young: $\sim$ 10--100 Myr.  By comparing HD 1160 B and C's derived luminosity to predictions from substellar cooling models given the system's age, N12 derived a mass of $24-45$~$M_{\rm J}$ for HD 1160 B and 0.18--0.25 $M_{\rm \odot}$ for HD 1106 C.

HD 1160 B was revisited as part of the first results from the Spectro-Polarimetric High-contrast Exoplanet REsearch \citep[SPHERE,][]{B08} at the Very Large Telescope (VLT) by \citet[][hereafter M16]{Maire15}. M16 observed HD 1160 B with VLT/SPHERE to obtain the first spectrum for the object, with resolution of $R\approx30$ covering the wavelength range of $1.0-1.6$~$\mu$m.  Based on HD 1160 B's spectral shape, M16 re-classify its spectral type as M6$^{+1}_{-0.5}$ and derive an effective temperature of $3000\pm100$ K but do not constrain the object's surface gravity their SPHERE $YJH$ spectrum.  Based largely on a wider range and systematically older age estimate ($30-300$ Myr), they estimate a mass of $39-144$~$M_{\rm J}$ based on bolometric luminosity and $69-172$~$M_{\rm J}$ based on effective temperature.

Two key questions surround the nature of HD 1160's companions.  The first is metallicity.  By comparing HD 1160 A's Hertzsprung-Russel (HR) diagram position to predictions from the \citet{Siess2000} isochrones, N12 found that the primary is only matched to isochrones with a substantially subsolar metallicity ([Fe/H] = $-0.3$, in conflict with the very red near-infrared colors of HD 1160 B and C.   While M16 derived bluer infrared colors for HD 1160 B and C, they found atmosphere models assuming a subsolar metallicity fit HD 1160 B's spectrophotometry the best. As most stars in the solar neighborhood have a near-solar metallicity, new analysis is needed to clarify whether or not HD 1160 A has a non-solar metallicity.  \textit{The atmospheric properties of HD 1160 B, including metallicity, could be clarified with new spectrophotometry with greater precision covering a wider wavelength baseline}.

The second question is the system's age.  Different age estimates for HD 1160 result in quite different inferred masses for the two low-mass companions.  HD 1160 B's interpretation changes from a low-mass, high mass ratio brown dwarf (for a young age) to an object straddling the brown dwarf/stellar boundary (for an older age).  \textit{The age of the HD 1160 system could be better clarified with a more in-depth and systematic analysis of possible age indicators for the primary and the two companions.}

In this work, we present high-precision photometry of the HD 1160 system obtained with the Subaru Coronagraphic Extreme Adaptive Optics (SCExAO) project in the $YJH$ passbands and integral field spectroscopy with the Gemini Planet Imager (GPI) in the $Y$ band (\S 2).  We compare HD 1160 B's spectrum and new photometry to that of young brown dwarfs and field objects and match its photometry/spectra using the BT-Settl atmosphere models (\S 3) to address whether or not HD 1160 B clearly shows evidence for a subsolar metallicity and derive an updated luminosity for both it and HD 1160 C.  We use an array of age estimates to better clarify the age of the system (\S 4).  After modeling HD 1160 B's atmosphere and better constraining HD 1160 C's luminosity, we use our updated age estimate to more precisely infer HD 1160 BC's mass (\S 5).  Section 6 summarizes our findings and identifies future work that can better clarify the properties of this system.


\section{Data}

\subsection{Subaru Coronagraphic Extreme Adaptive Optics $YJH$ Imaging\label{sec:scexaoyjh}}

The Subaru Coronagraphic Extreme Adaptive Optics (SCExAO) is a high contrast imaging platform designed for the discovery and characterization of faint companions \citep{scexao15}. A partial correction of the low order modes of the wavefront is performed by the Subaru facility adaptive optics instrument (AO188), and a final correction, including the high-order modes is performed by a visible-light pyramid wavefront sensor and a 2000-element deformable mirror inside SCExAO. Ultimately, the project is slated to yield 90+\% Strehl ratios at 1.6 $\mu$m on bright stars (I (mag) $<$ 9--10).  

On 31 October 2015, we observed HD 1160 with SCExAO coupled with the near infrared HiCIAO camera \citep{HICIAO09} in the $Y$, $J$, and $H$ broadband filters ($\sim$ $8.0$ mas pixel$^{-1}$)\footnote{The plate scale when using AO188 and HiCIAO is $9.5$ mas pixel$^{-1}$, but for SCExAO and HiCIAO, the plate scale is $8.3$ mas pixel$^{-1}$.}.  All observations were obtained in angular differential imaging (ADI) mode \citep{Marois06} but with a very small amount of field rotation ($\Delta$P.A. $\sim$ 1.5--2 $^{o}$). 

While our instantaneous Strehl ratios were in the 80--90\% range, strong low frequency vibrations induced by the telescope drive system degraded AO correction quality to $\sim$ 60--70\% Strehl over the course of $>$ 0.1s observations.  These vibrations are now partially mitigated by a linear-quadratic-gaussian (LQG) controller, yielding slightly improved performance \citep{Lozi2016}.  Upcoming improvements -- using the higher-stroke AO188 deformable mirror driven by multiple accelerometers to offload tip-tilt variations -- should eliminate these vibrations.  
Nevertheless, the partial correction offered by SCExAO substantially improves upon the image quality achieved with discovery images of HD 1160 B obtained in the near-infrared (NIR) from Gemini/NICI.    

Our observing log is detailed in Table~\ref{table:obslog}.  We cycled through the three near-infrared passbands using a combination of short, lightly saturated exposures and unsaturated images obtained using HiCIAO's neutral density filters. Basic image processing followed standard methods employed in \citet{Currie11,Currie14a} for flat fielding, dark subtraction, de-striping, radial profile subtraction, and combining images. As HD 1160 B/C were plainly visible in the raw data, we did not perform any point-spread function (PSF) subtraction but instead simply removed the azimuthally-averaged seeing halo.  

\begin{figure}[ht]
\begin{center}
\includegraphics[scale=0.8]{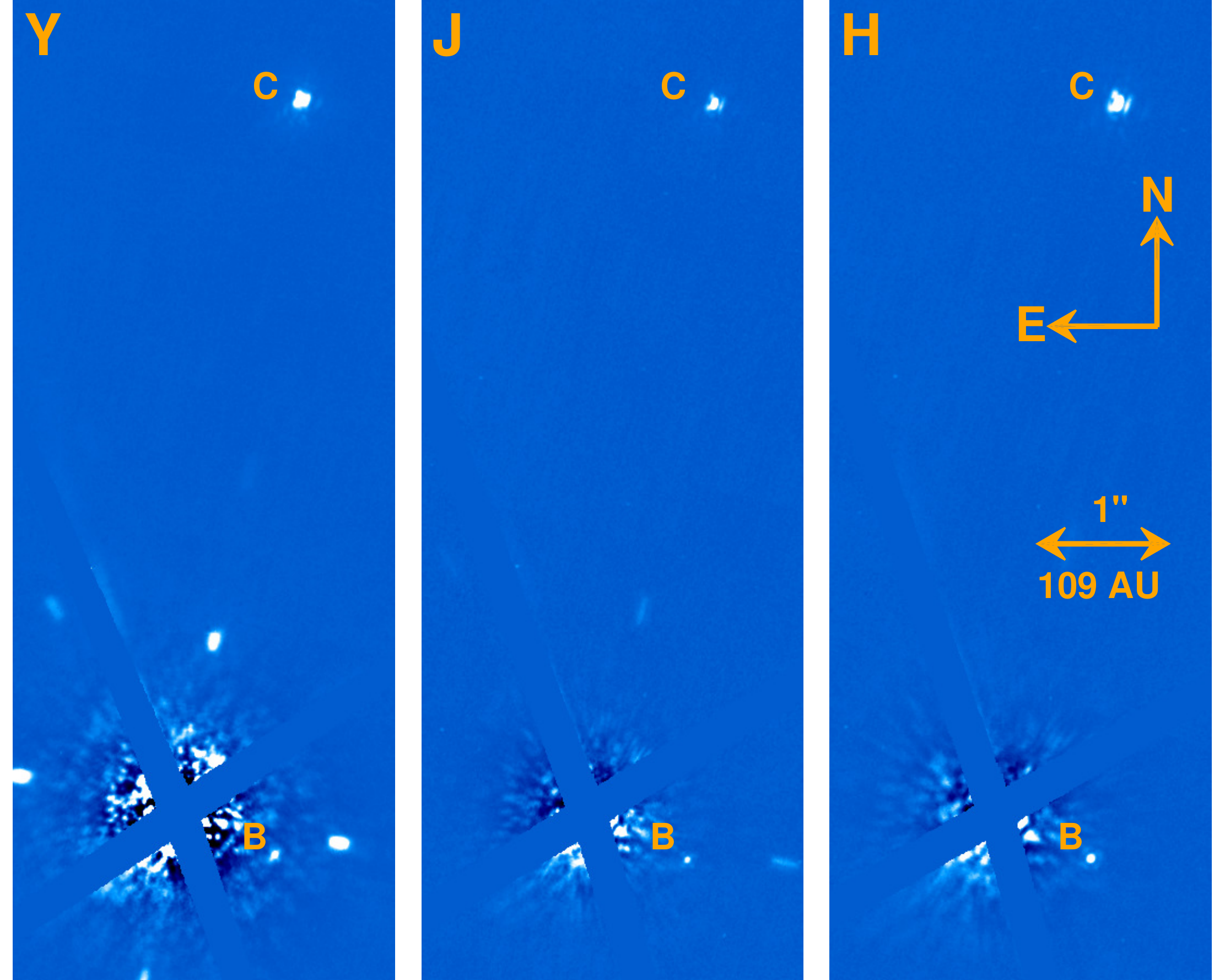} 
\end{center}
\caption{\label{fig:scexaoimage} SCExAO observations of HD 1160 in $Y$ (left), $J$ (middle), and $H$ (right).  Diffracted light from the secondary spider is masked in each panel.  The four bright, elongated speckles 2\arcsec{} from the star are scattered light from SCExAO, the result of quilting from the deformable mirror, which can be used to aid image registration.  The companions HD 1160 B and C are detected at a high SNR in each image.  
}
\end{figure}

Despite the short integration times and no PSF subtraction techniques applied, both HD 1160 B and C were easily recovered with a signal-to-noise ratio (SNR) of $\sim$ 70 and $>200$, respectively, in $YJH$ passbands (Figure \ref{fig:scexaoimage}).  While processing the data with advanced PSF subtraction techniques like A-LOCI or KLIP \citep{Currie15a,Soummer12} yielded stronger detections and lower residuals at smaller separations, they also partially attenuate the companion signal.  For simplicity, we analyze only the halo-subtracted images which lack companion signal loss.

Our SCExAO photometry of HD 1160 B are shown in the left columns of Table~\ref{table:phot}:
m($Y$,$J$,$H$)=\scexaoyb,~\scexaojb, and~\scexaohb. For HD 1160 C, we derive m($Y$,$J$,$H$)=\scexaoyc~, \scexaojc, and~\scexaohc. Our errors consider (in quadrature) the SNR of the detection (computed as in \citealt{Currie15a}), uncertainties in the primary magnitude, and a minor contribution of uncertainties in the attenuation of the neutral density filter.  

We use aperture photometry to measure the brightness of HD 1160 A (in unsaturated images obtained with the neutral density filter) and HD 1160 BC.  For the fainter HD 1160 B companion, we derived the contrast between it and the primary using a range of aperture diameters comparable to the FWHM of a point source in each passband.   For the contrast of the brighter, wider-separation HD 1160 C component, we used larger apertures enclosing much more ($>$ 90\%) of the primary and companion signals to guard against flux loss due to PSF degradation (e.g. vignetting of the field).   

Our SCExAO HD 1160 B $J$ and $H$ photometry shows good agreement with SPHERE measurements (Table~\ref{table:phot}, third column). SPHERE's brightness estimates are systematically brighter by 1.8$\sigma$ ($0.22\pm0.12$ mag) at $J$-band and $\approx2.8\sigma$ ($0.23\pm0.08$ mag) at $H$-band. A good portion of the discrepancy at $H$-band is due to SPHERE's spectrum only covering half the $H$-band filter.  Small differences at $J$ band may be due to SPHERE's flux calibration.  SPHERE's spectro-photometry had to be corrected for signal loss due to aggressive processing; while the HiCIAO neutral density filter is flat across the bandpass, SPHERE's exhibits a slight wavelength dependent attenuation at $J$ band.  

In contrast, our photometry strongly disagrees with the measurements of~N12~by 1.14 mag at $J$ and 0.44 at $H$  (Table~\ref{table:phot}, columns three and six)\footnote{For another estimate of HD 1160 B's $J$-band photometry, we downloaded and reduced archival $J$-band Keck/NIRC2 data for HD 1160 from 2004-07-14 (P.I. I. Song) using image processing methods identical to those described in this work.  Our Keck detection has a substantially lower SNR than SCExAO and SPHERE.   However, it agrees with the SCExAO and SPHERE photometry against the NICI photometry: m$_{J}$ = 14.54 $\pm$ 0.24.}. Similar discrepancies between NICI photometry and other instruments \citep{Bocca13} have been observed previously. The difference may be due to inaccurate flux normalization, possibly from mismeasured attenuation through the coronagraphic mask, or an inaccurate subtraction of the sky background annulus surrounding the companion \citep[as was the case for early photometry of ROXs 42B``c", see][]{Currie14b}. 


\subsection{Gemini Planet Imager $Y$ band Low-Resolution Spectroscopy}
\subsubsection{Observations and Basic Processing}
HD 1160 B was observed with the \textit{Gemini Planet Imager} \citep{Macintosh06,Mac14com} (GPI) in $Y$ band (\gpiwave~$\mu$m, $R\approx37$) on
17 November 2013 in $\sim$ 1\arcsec{} seeing by the GPI Verification and Commissioning team (Table~\ref{table:obslog}). The observations consist of nine 88.7 s exposures in IFS mode \citep{Larkin14} using GPI's apodized pupil Lyot coronagraph with a pixel scale of $14.14\pm0.01$ mas pixel$^{-1}$ \citep{Konopacky14}.  As shown in Figure \ref{fig:filt}, the GPI $Y$ band filter extends 0.1~$\mu$m past typical $Y$ band filters \citep[See the appendix of][]{Liu12}.  It also extends blueward of the SPHERE $Y$ band data from M16, thus expanding the wavelength range at which HD 1160 B can be detected and analyzed beyond that offered by either SPHERE or SCExAO observations.  The observations were obtained in ADI mode.  However, like the SCExAO observations, the GPI sequence covered little parallactic motion ($\Delta$P.A. $\sim$ 4.3$^{o}$).  

We processed the images using the GPI data reduction
pipeline version 1.3.0 \citep{Maire10,Maire12,Perrin14}. The pipeline requires the location and spectral solution for every lenslet on the HAWAII-2RG detector. These lenslet locations
were determined by using a cross correlation between 
the deep argon calibration source images available on the 
GPI public webpage\footnote{http://www.gemini.edu/sciops/instruments/gpi/public-data}.
The elevation of the telescope differed between
the images of HD 1160 B and the daytime argon calibration lamp sequence. We calculated a shift to determine the overall change of the
wavelength solution between the daytime calibrations and that
appropriate for the observations of HD 1160 B. 

We further used the GPI data reduction pipeline to apply dark corrections,
remove bad pixels, track satellite spot locations,
convert each microspectra into a 37 channel spectral cube
(\gpiwave~$\mu$m), correct for distortion using the publicly available distortion solution$^1$, and correct for atmospheric differential refraction. Each of the 9 individual raw observations were processed in an identical way. HD 1160 B is visible in individual spectral channels of each data cube (Figure \ref{fig:avg}).

\subsubsection{Extraction of HD 1160 B Spectrum and Calibration\label{sec:gpispec}}

\begin{figure}[ht]
\begin{center}
\includegraphics[trim={0 3.4cm 0 3.4cm},clip,width=\textwidth]{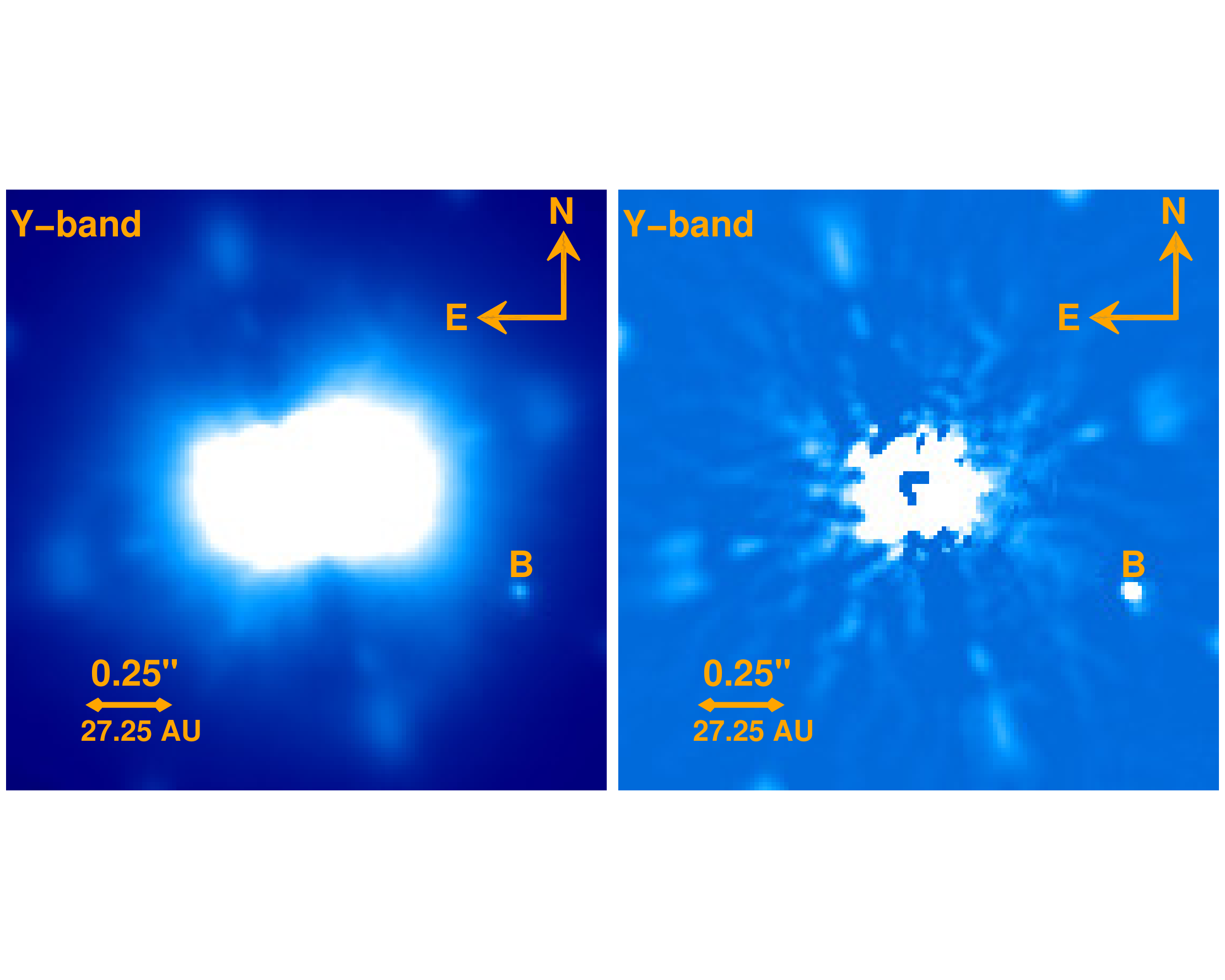}  
\end{center}
\caption{\label{fig:avg} The wavelength-collapsed, de-rotated, stack of temporal frames for Gemini Planet Imager $Y$ band observations of HD 1160 B. To remove the slowly varying background due to the smooth speckle halo, the original image (left) is high-pass filtered (right) using a median with a box size of 11 pixels ($\sim5\lambda/D$). 
}
\end{figure}

The final, distortion-corrected, de-rotated, time-averaged data cube of HD 1160 has a smooth, slowly-varying halo at HD 1160 B's angular separation (left panel, Figure~\ref{fig:avg}).  We remove this slow-varying background using a median high pass filter (right panel, Figure~\ref{fig:avg}) with a box size of 11 pixels $(\approx5\lambda/D)$. 
We manually confirmed that this filtering did not result in a measurable loss in signal. Our final data cube has an SNR $\sim$ 90 detection for HD 1160 B in the wavelength-collapsed cube and SNR $>$ 10 for each channel. 

To flux-calibrate our final data cube and extract HD 1160 B's flux-calibrated spectrum, we again followed steps outlined in the GPI data reduction pipeline.  First, we extracted the flux-calibrated spectra of the four satellite spots using the laboratory measured satellite-to-star flux ratio of \gpiysattostar~from~\cite{Wang14} and assumed an A0V template spectrum from the Pickles stellar library \citep{Pickles98} to characterize the host star spectrum of HD~1160~A (N12) \footnote{HD~1160~A is a well-known IR photometric standard star \citep{Elias82}.}.  We adopted a 2-pixel aperture radius,  a 5--20 pixel annulus to define the sky background, and the standard deviation of the four satellite spot flux density measurements as the uncertainty in absolute calibration in each spectral channel\footnote{The upper-left spot spectrum contributed the most to the uncertainty in each spectral channel and in the wavelength-collapsed image, which is $\sim$ 20\% fainter than the average of the other three spots.}.   

\begin{figure}[ht]
\begin{center}
\includegraphics[angle=90,width=\textwidth]{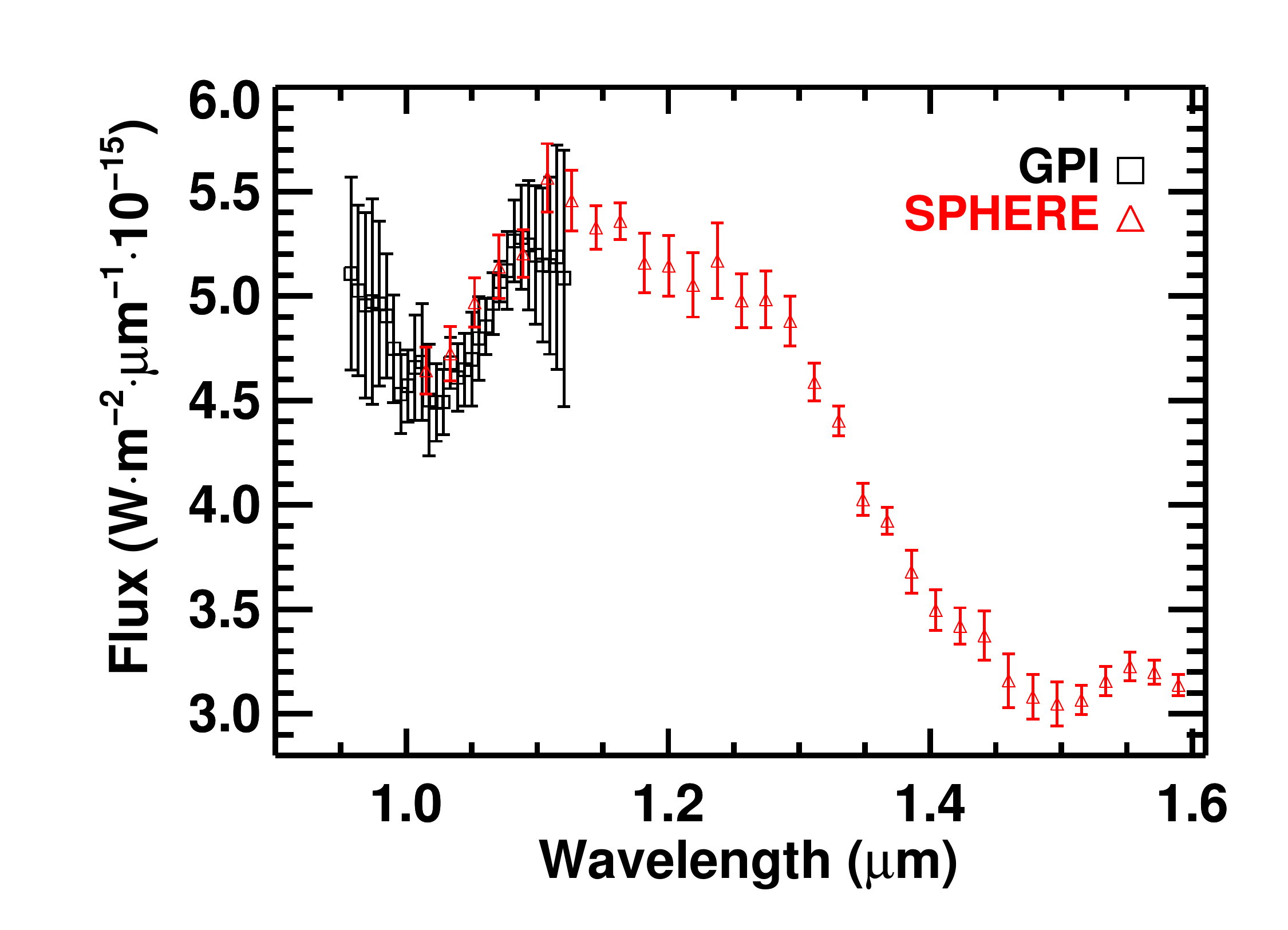}  
\end{center}
\caption{\label{fig:spec} Comparison between the GPI $Y$ band spectrum of HD 1160 B (\gpiwave~$\mu$m, black squares) and the SPHERE spectrum (red triangles) covering $1.0-1.6$~$\mu$m.
There is good agreement between the SPHERE spectra and our higher resolution GPI  spectrum (see~\S\ref{sec:gpispec}) within errors, although the SPHERE spectra are slightly brighter over the $Y$ bandpass.   The GPI spectra errors are larger due to the higher resolution (the wavelength bins are $\sim5.4$~nm for GPI and $\sim18.5$~nm for SPHERE). We use a combined GPI and SPHERE spectrum for our atmosphere modeling (see~\ref{sec:atmos}). 
}
\end{figure}

To extract the spectrum of HD 1160 B, we used the same aperture radius and background annulus used to extract the satellite spot spectra.  The errors in the extracted and flux-calibrated HD 1160 B spectrum draw from the uncertainty in the spot calibration and the uncertainty in the background annulus signal surrounding HD 1160 B\footnote{For a separate spectral extraction independent of the GPI data reduction pipeline, we performed aperture photometry using \textrm{aper.pro}\footnote{http://idlastro.gsfc.nasa.gov/ftp/pro/idlphot/aper.pro} at the location of the companion, for each wavelength slice, assuming the background to be zero.  The two methods show strong agreement, to within the calculated errors/channel.}.  We trimmed the HD 1160 B spectrum of the first and last few channels ($<$\gpishortcutoff~$\mu$m~and~$>$\gpilongcutoff~$\mu$m), which had absolute calibration uncertainties $\gtrsim$ 10\%.  The resulting flux calibrated spectrum spanning~\gpiwave~$\mu$m is shown in Figure~\ref{fig:spec}.   From the wavelength-collapsed data cube, we estimate a $Y$ band magnitude of m($Y$)$=15.13\pm0.21$, consistent with our more precise SCExAO measurements (Table~\ref{table:phot}, third column).

\subsection{Rereduced Keck/NIRC2 $M^\prime$ Photometry\label{sec:keckm}}
M16 found significant discrepancies between their SPHERE/IRDIS photometry and that from N12 in the $J$ and $H$ passbands.  Additionally, even after adopting the revised HD 1160 B NaCo $L^\prime$ photometry from M16 ($\sim$ 0.2 mag fainter), the implied $L^\prime-M^\prime$ color is $\sim -0.7$, characteristic of L/T transition objects but far too blue for a mid M dwarf like HD 1160 B \citep[see ][]{Galicher11}.  

Therefore, we rereduced the same archival Keck/NIRC2 $M^\prime$ data first analyzed by N12 (HD 1160 C was outside of the field of view).  Basic processing followed standard steps we have previously used for thermal IR data \citep[e.g.][]{Currie11} but without PSF subtraction.  Briefly, after applying a linearity correction to the data, we constructed a sky frame for a given image from the 5 nearest (in time) images where the star is in a different dither position.  After sky subtraction, we corrected for distortion, registered each image to a common center, subtracted off the radial intensity profile for each image, averaged the set of profile-subtracted images using 3-$\sigma$ outlier rejection, and applied a 5 $\lambda$/D moving-box median filter to the combined image.  As the combined image shows some residuals of the PSF halo, we computed the SNR of the detection in a conservative fashion (not assuming that the image at HD 1160 B's location is photon noise dominated) as we have typically done with high-contrast imaging data sets \citep[e.g.][]{Currie11}.

Our re-reduction yields a SNR $\sim$ 4.6 detection of HD 1160 B with a measured apparent magnitude of m($M^\prime$) = 13.81$\pm$ 0.24 (Table~\ref{table:phot}, 
far-right column), where the intrinsic SNR of the detection contributes almost all of the photometric uncertainty.  HD 1160 B is about 63\% brighter at $M^\prime$ than reported in N12.  The implied $L^\prime$-$M^\prime$ color of $\sim$ -0.21$\pm$0.23 is in agreement with expected colors for mid M to early L objects \citep{Galicher11}.

\section{Characterization of HD 1160 B and C}

\subsection{Analysis of the HD 1160 B and C Photometry\label{sec:phot}}

\begin{figure}[ht]
       \centering
           \includegraphics[trim={0.8cm 0.8cm 0.8cm 0.8cm},clip,angle=90,scale=0.38]{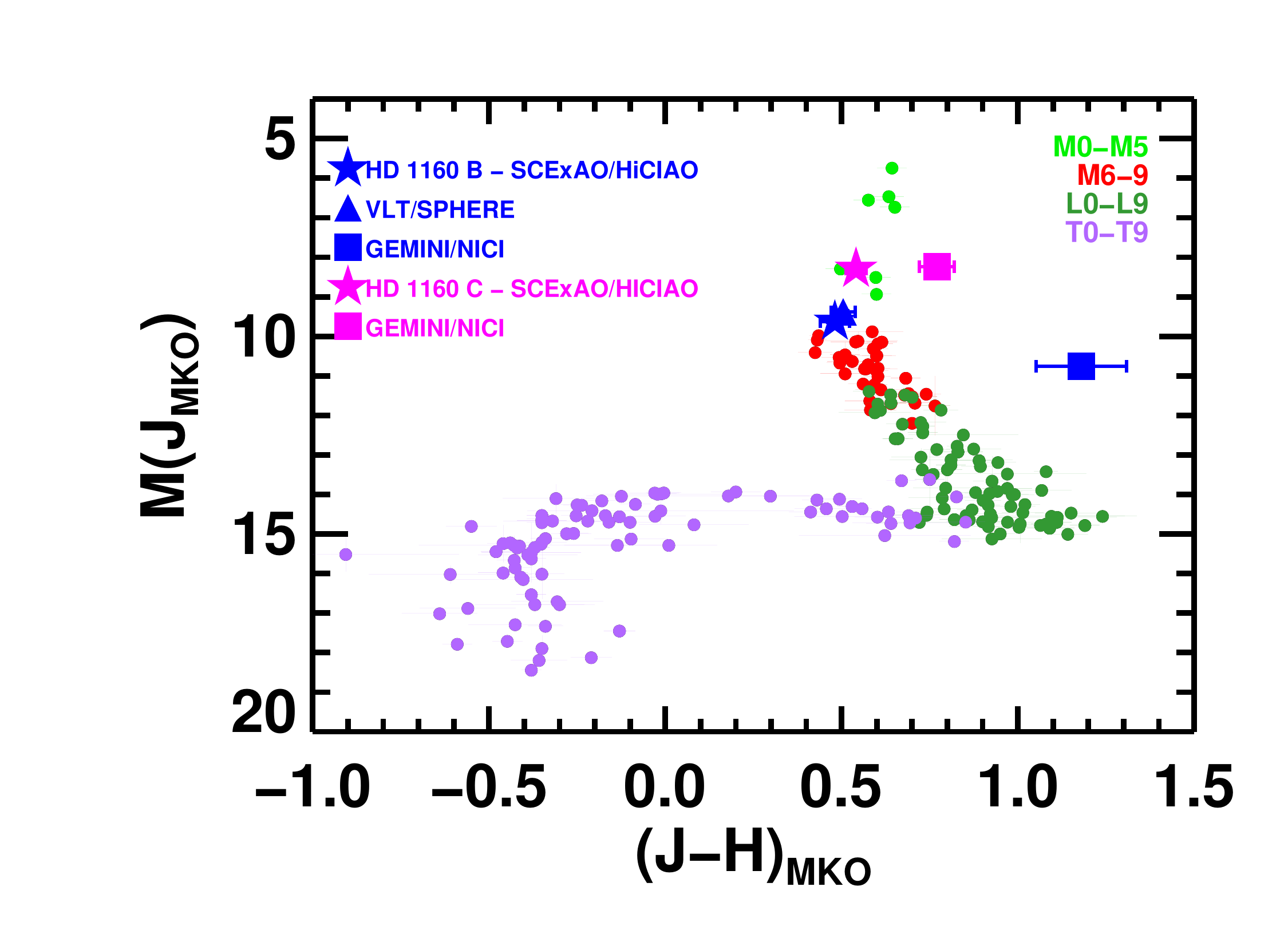}
            \includegraphics[trim={0.8cm 0.8cm 0.8cm 0.8cm},clip,angle=90,scale=0.38]{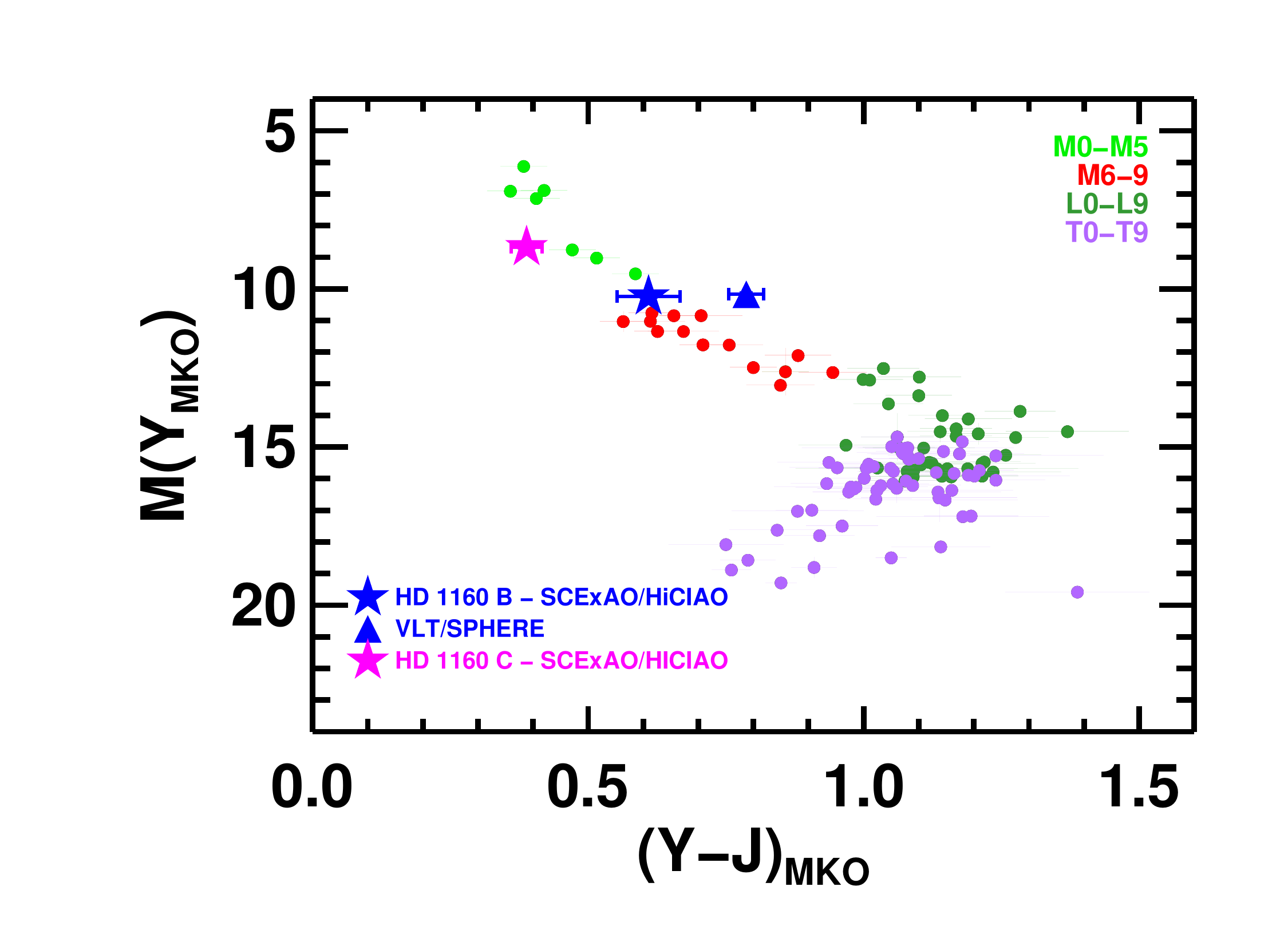}
    \caption{
        Color-Magnitude diagrams comparing SCExAO/HiCIAO (this work), Gemini/NICI photometry \citep{Nielsen12}, and the integrated-spectra photometry of SPHERE observations \citep{Maire15}. The M3--M6 main sequence is from dwarf standards of \cite{Kirkpatrick10}. The M6--T9 main-sequence dwarfs that form the color-magnitude sequence is adopted from \cite{Dupuy12}. At $Y$, \cite{Dupuy12} integrated spectra for the M6--T9 dwarfs to compute the photometry using the UKIRT/WFCAM $Y$ band 
        \citep{Hewett06} 
        .
        }
        \label{fig:cmd}
    \end{figure}

To compare HD 1160 B and C to field MLT dwarfs, we adopt our SCExAO $YJH$ photometry. We assume a distance of~\distance~pc \citep{vanLeeuwen07}, yielding absolute magnitudes of~$Y$=\absmagy,~$J$=\absmagj~and~$H$=\absmagh~mag. Figure~\ref{fig:cmd} compares our HD 1160 B photometry to the sequence of MLT dwarfs from \cite{Dupuy12}, and M-dwarf standards of \cite{Kirkpatrick10} and references therein. We plot $YJH$ colors of $J-H$=\jhcolor,~ $Y-J$=\yjcolor~colors and the absolute magnitudes in $Y$, $J$, passbands for HD 1160 B as the blue star in Figure~\ref{fig:cmd}. In all cases, HD 1160 B's photometric points are consistent with the field dwarfs sequence.  We find HD 1160 B's colors in agreement with the M5$-$M6 spectral standards of \cite{Kirkpatrick10} (transition of green and red diamonds in Figure~\ref{fig:cmd}). Similarly, as shown by the pink star in Figure~\ref{fig:cmd}, we find the 
near-IR colors and the $YJH$ absolute magnitudes of HD 1160 C in agreement with M3$-$M4 spectral standards of \cite{Kirkpatrick10}. 

As a further check, we compare HD 1160 B's nominal colors to those if we adopt photometry extracted from the SPHERE spectrum in $JH$ (Table~\ref{table:phot}) and photometry from partial coverage in $Y$ band (m($Y$)~=~\vltyb; M($Y$)=$10.12\pm0.02$).  The SPHERE-derived photometry is slightly redder in $Y-J$ but otherwise suggests that HD 1160 B is consistent with the field sequence (pink and blue triangles, Figure~\ref{fig:cmd}).

\subsection{Analysis of the HD 1160 B Spectrum \label{sec:spectype}}

To compare HD 1160 B's full near-infrared spectrum to that from other objects, we construct a merged $YJH$ spectrum from GPI and SPHERE data.   We applied an offset to the GPI spectrum such that its photometry integrated over $Y$ band perfectly matched the SCExAO $Y$ band photometry and a separate offset to the SPHERE spectrum such that its band-integrated photometry matched that of SCExAO's as well.  We adopt the GPI (SPHERE) measurements shortwards (longwards) of 1.12 $\mu m$.  

As implied by N12 and M16 and shown in \S\ref{sec:hd1160age} of this paper, HD 1160's age is likely intermediate between that of low mass members of young star forming regions such as Upper Sco \citep[$11\pm2$ Myr,][]{Pecaut12} and older field objects ($>1$ Gyr). For spectral typing HD 1160 B, we therefore compare the GPI+SPHERE spectrum to field M dwarf spectral standards from the SPEX library \citep{Rayner03} and Upper Sco members \citep{Dawson14}.   

We perform the spectral typing by binning the template NIR spectra to the resolution of the GPI and SPHERE spectra and computing the $\chi^{2}$ as: 
\begin{equation} 
\chi^{2} = \sum_{i=1}^{i=57}\frac{(F_{\rm i,obs} - F_{\rm i,template})^{2}}{\sigma_{\rm i,obs}^{2}+\sigma_{\rm i,template}^{2}}
\end{equation}
where flux $F$ is in watts~m$^{-2}$~$\mu m^{-1}$ and index $i$ corresponds to one of the 57 spectral channels of the flux calibrated GPI and SPHERE spectra. We weighted the GPI and SPHERE spectra equally.  

\begin{figure}[ht]
\begin{center}
\includegraphics[angle=90,width=\textwidth]{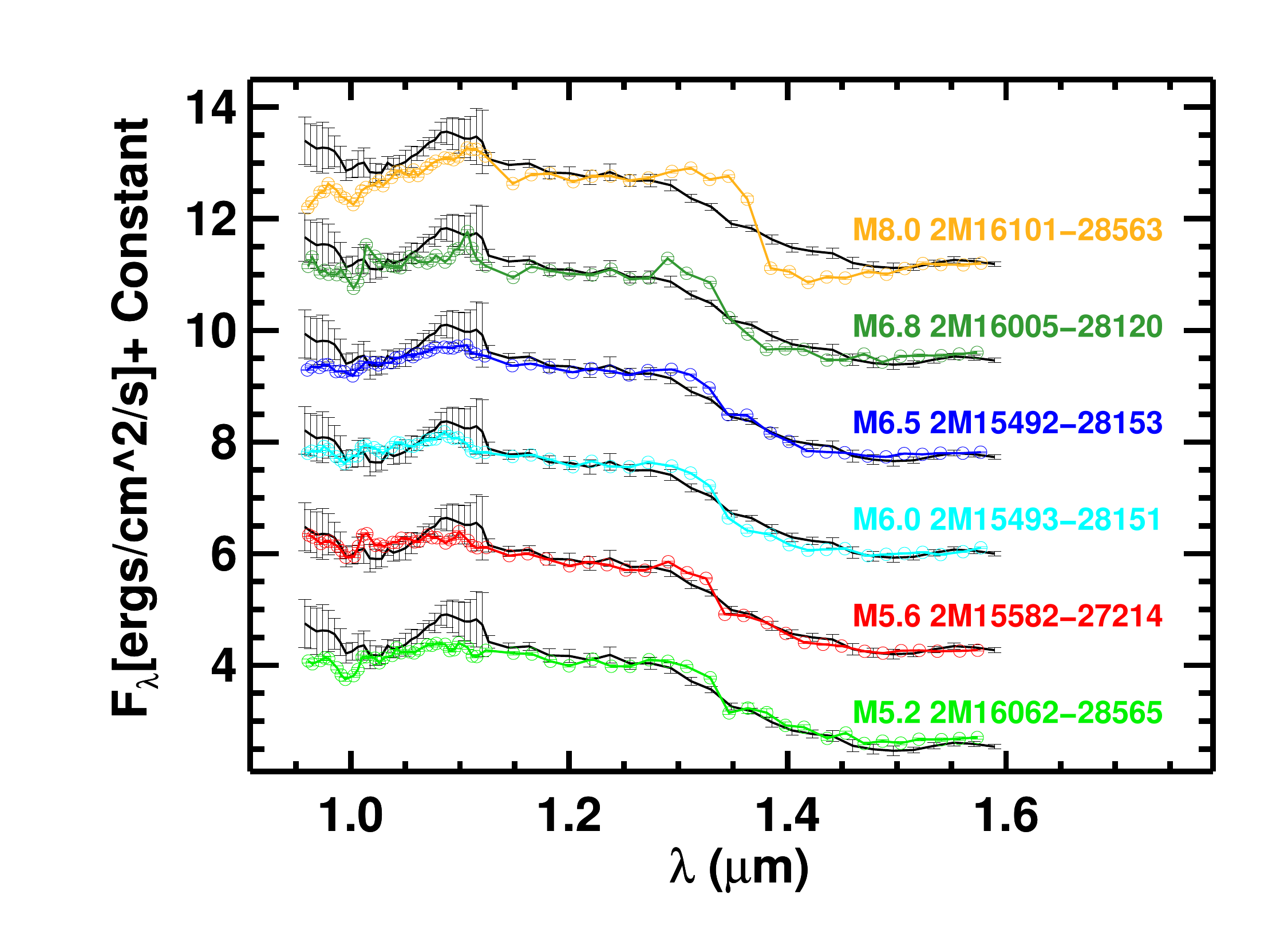}  
\end{center}
\caption{\label{fig:uppersco}
The combined SPHERE and GPI spectra of HD 1160 B are best reproduced by the SPEX spectra of M5.2-M6.8 members of Upper Sco from \cite{Dawson14}. M8 and later members of Upper Sco do not reproduce the semi-smooth slope of the SPHERE spectrum from 1.2-1.6~$\mu$m. The SPEX spectra were binned to the resolution of the GPI and SPHERE respectively, and scaled to the corresponding flux levels to minimize the $\chi^{2}$. Given that the GPI spectrum is higher resolution and the SPHERE spectrum has greater wavelength coverage, we weighted both equally in our spectral typing procedure (see \S\ref{sec:spectype}). $F_{\lambda}$ is normalized by a factor of $10^{-15}$. The constant is an an offset in unit steps of $1.7\times10^{-15}$ ergs/cm$^{2}$/s. 
}
\end{figure}

As shown in Figure~\ref{fig:uppersco}, the GPI and SPHERE spectrum shape is poorly reproduced for Upper Sco M8 member 2M16101$-$28563 for spectral channels $<1.0$~$\mu$m and 1.3$-$1.4~$\mu$m. The disagreement is greater for L0.8 Upper Sco member 2M16195$-$28322 (top of Figure~\ref{fig:uppersco}). On the other hand, M$5.2-$M$6.8$ Upper Sco members have $\Delta\chi^{2}$ within the $95\%$ confidence interval for 57 degrees of freedom \citep[see Numerical Recipes,][]{Press92}. We therefore find the spectral type of HD 1160 B as compared to Upper Sco Members to be M$6^{+0.8}_{-0.8}$. 

\begin{figure}[ht]
\begin{center}
\includegraphics[angle=90,width=\textwidth]{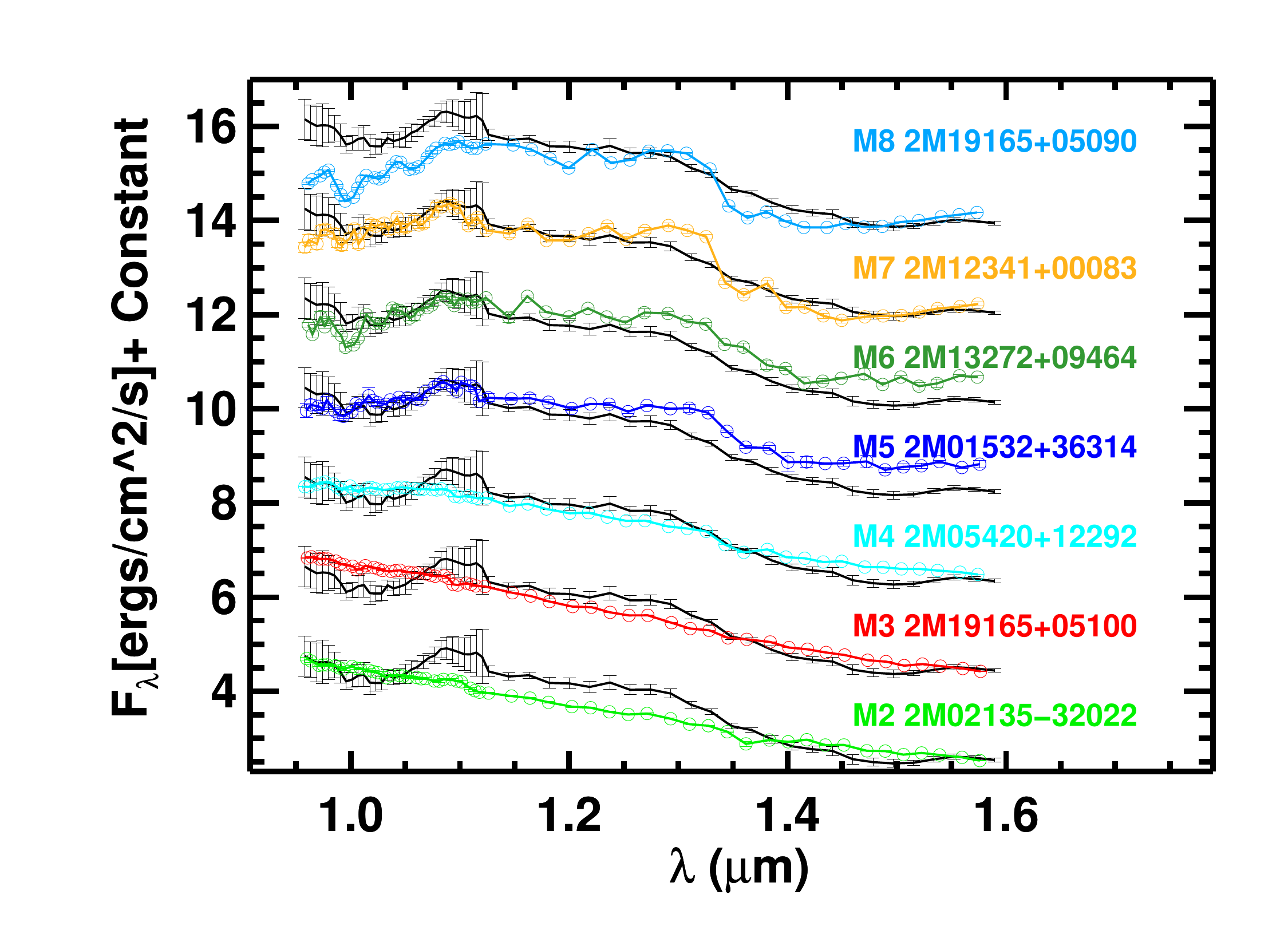}  
\end{center}
\caption{\label{fig:thefield}
Similar to figure~\ref{fig:uppersco}, the combined SPHERE and GPI spectra of HD 1160 B are best reproduced by the SPEX spectra of M5--M6 field standards from \cite{Kirkpatrick10} (c.f.\ Figure 9). Standards M4 or earlier spectra do not reproduce the overall shape of the HD 1160 spectrum, while standards M7 and later also do not accurately produce the shape of the HD 1160 spectrum from 1.3$-$1.6~$\mu$m. The method of comparing standard spectral templates to the spectrum of HD 1160 is identical to figure~\ref{fig:uppersco} and detailed in \S\ref{sec:spectype}. $F_{\lambda}$ is    ed by a factor of $10^{-15}$. The constant is an an offset in unit steps of $1.9\times10^{-15}$ ergs/cm$^{2}$/s. 
}
\end{figure}

Similarly, Figure~\ref{fig:thefield} compares the GPI and SPHERE spectra of HD 1160 B to M2$-$M8 field SPEX spectra standards from \cite{Kirkpatrick10}. We find that the overall spectral shape is best reproduced by M5 standard 2M01532+36314 or M6 standard 2M13272+09464. The overall shape of the spectrum is poorly reproduced by the linear slope of M4 or earlier standards across $0.95-1.6$~$\mu$m. Standard spectra later than M6 poorly reproduce the smooth portion of the HD 1160 B spectrum at $1.3$--$1.4$~$\mu$m as well as the slope at $<1.0$~$\mu$m. Therefore, we find the spectral type of HD 1160 B as compared to field M dwarfs to be~M$5.5^{+0.5}_{-0.5}$,~in agreement with the spectral types derived for HD 1160 B compared to Upper Sco members above. We adopt a spectral type for HD 1160 B of~\sptb~as an average between the dwarf and Upper Sco spectral types, finding agreement with the M16 spectral type of M$6_{-0.5}^{+1.0}$. 


\subsection{Atmospheric Modeling of HD 1160 B \label{sec:atmos}}

{\it Atmosphere Models Considered.} To explore HD 1160 B's atmospheric properties, we compare the object’s broadband photometry and $YJH$ GPI+SPHERE spectrum to predictions from planet atmosphere models, adopting a range of effective temperatures and surface gravities. We consider the BT-Settl 2013 models of \cite{Allard12} with temperatures of $2500-3500$~K, surface gravities of log $g=3.5-5.5$ dex and solar metallicity. Using the same atmosphere models, M16 concluded that spectral energy distribution (SED) modeling favors a subsolar metallicity for HD 1160 B.  Using revised/new photometry and spectroscopy, we assess whether or not HD 1160 B may yet have a solar metallicity.


{\it Fitting Method.} Following previous approaches \citep[e.g.][]{Currie14b}, we fit our $YJH$ (SCExAO, this work) 
$K_{s}$ (SPHERE, M16) $L^{\prime}$ (NaCo, M16) $M^{\prime}$ 
(Keck, this work) photometry and $YJH$ GPI+SPHERE spectroscopy.  To model photometry, we compare the measured flux densities with predicted ones from the model spectra convolved with the MKO filter functions. The BT-Settl grid has significantly higher spectral resolution than the data (R$\approx$37, SNR$>>10$ per channel). Thus, to 
model the spectrum of HD 1160 B, we rebin the BT-Settl spectra to the resolution of our extracted spectrum. We treat the planet radius as a free parameter, varying it from $R$ $\sim$ 0.75 $R_{\rm J}$ to $R$ $\sim$ 2.25 $R_{\rm J}$, which spans the wide range in radii predicted from models \citep{Baraffe03}. 

We identify the set of BT-Settl models consistent with the spectra or photometry at the 95\% confidence limits given the number of degrees of freedom. We adopt the measured photometric error, listed in Table~\ref{table:phot}, for model fitting.  We incorporate our distance uncertainty of $\pm 5.3$ pc \citep{vanLeeuwen07} in the uncertainty estimate for the planet radius.  We do not fit for $T_{\rm eff}$ and log $g$ parameters via $\chi^{2}$ minimization; rather, we compare the $\chi^{2}$ to a grid of atmosphere models with pre-determined $T_{\rm eff}$ and log $g$, treating the radius as a scaling parameter. Given six photometric points and four degrees of freedom, the 95\% cutoff is $\chi^{2}_{lim}$ = 9.49.   Given 57 spectral channels and 55 degrees of freedom, models with $\chi^{2}$ $\le$ 73.3 are consistent with the data. 

\begin{figure}[ht]
\begin{center}
\includegraphics[angle=90,width=\textwidth]{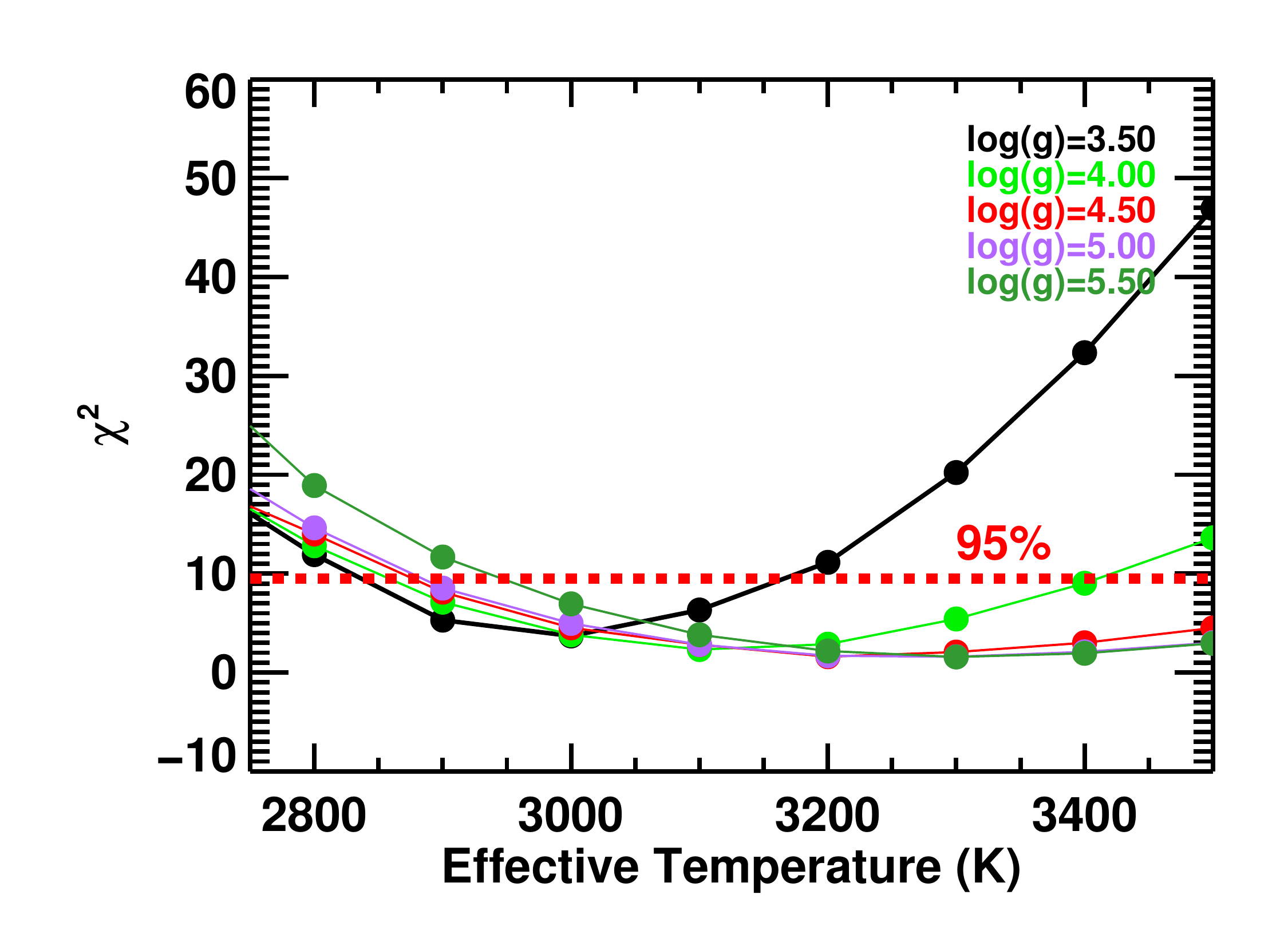}  
\end{center}
\caption{\label{fig:chi2phot} We find the photometry unable to constrain HD 1160 B's effective temperature to better than~$T_{\rm eff}$=\teffbestphot~$K$ and surface gravity $\log{(g)}$ to~\loggbestphot~dex. $\chi^{2}$ distributions for fitting HD 1160 B photometry as a function effective temperature and surface gravity. The horizontal dashed red line corresponds to the 95\% ($2\sigma$) confidence limit, identifying acceptably fitting models. Data points (circles) below this line represent BT-settl models that are in good agreement with the photometry of HD 1160 B. 
}
\end{figure}
\begin{figure}[ht]
\begin{center}
\includegraphics[angle=90,width=\textwidth]{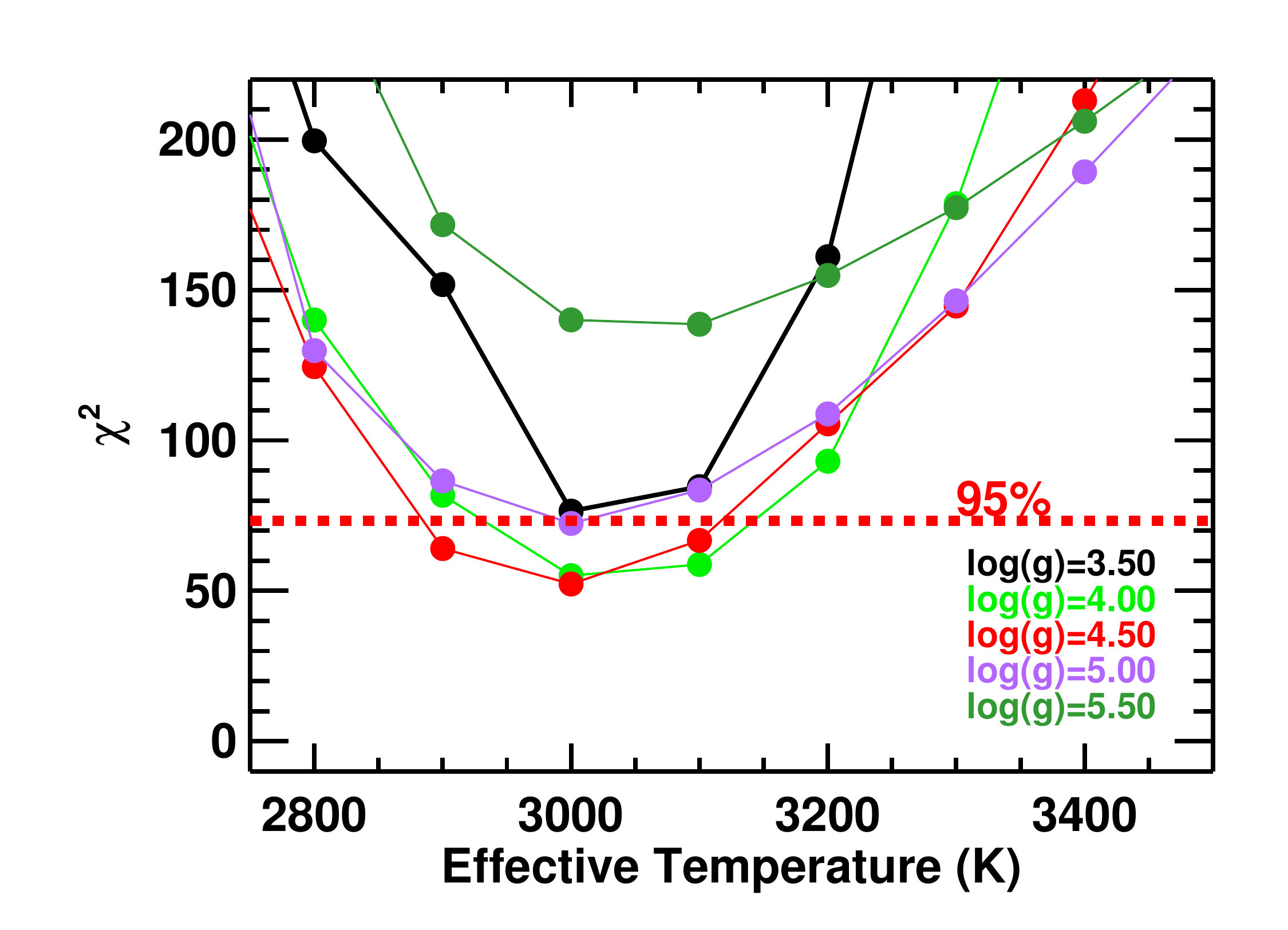}  
\end{center}
\caption{\label{fig:chi2spec} Same as Figure~\ref{fig:chi2phot} except 
for the spectroscopy. With the $0.95-1.6$~$\mu$m spectroscopy we find the best fits for 
an effective temperature~$T_{\rm eff}=$\teffb~$K$ and surface gravity $log{(g)}=$\loggb~dex, while $T_{\rm eff} = 2900$~$K$, log $g$ = 4.5 and $T_{\rm eff} = 3000$~$K$, log $g$ = 5.0 are also marginally acceptable.
}
\end{figure}
\begin{figure}[ht]
        \centering
        \includegraphics[scale=0.45]{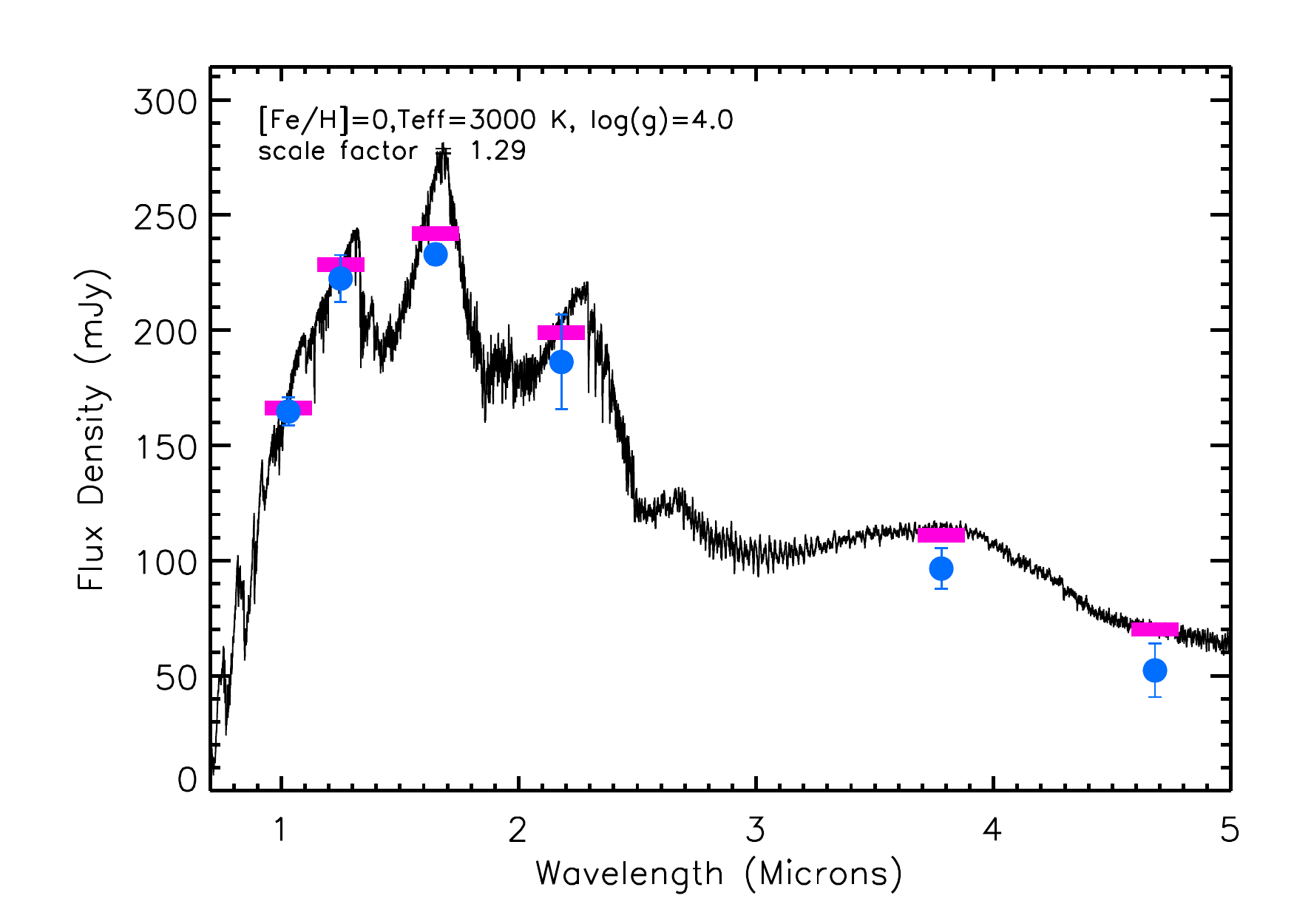}
         \includegraphics[scale=0.45]{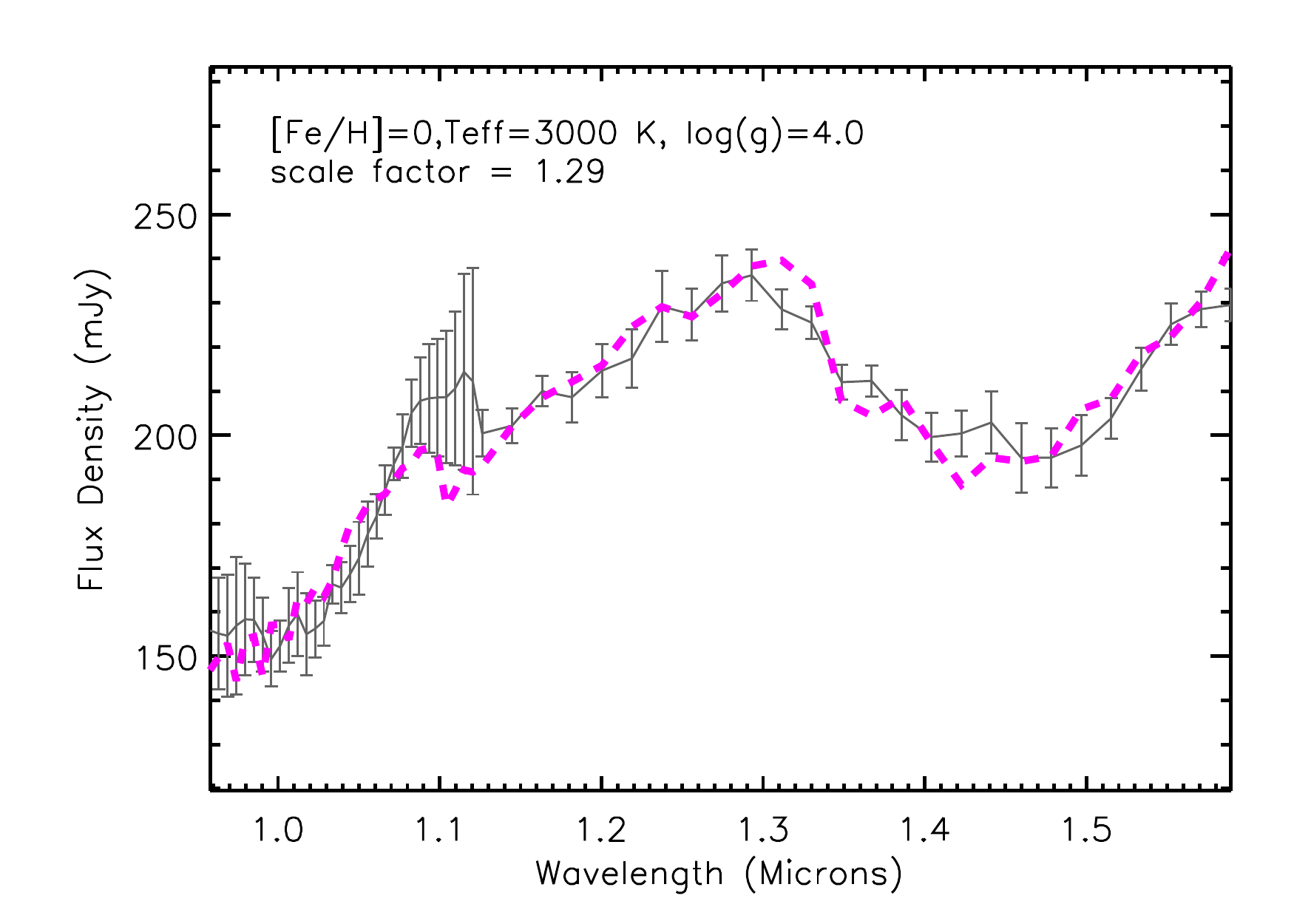}
        \caption[ SED fits]
        {\label{fig:sed}
        Example model fitting both the photometry (left) and spectroscopy (right) of HD 1160 B: $T_{\rm eff}$ = 3000~$K$, log $g$ = 4.0.  The photometric points are denoted by gray dots. In the left panel, the model is depicted as a thick black line and horizontal magenta lines shows the predicted photometry from the model in photometric bandpasses.  In the right panel, the spectrum is denoted by a grey line, and the model is depicted as a magenta dashed line. 
        }

\end{figure}

{\it Photometry and Spectroscopy Fitting Results.} Table~\ref{table:sed} summarizes our model-fitting results to the  $YJHK_{s}L^{\prime}M^{\prime}$ 
photometry and $0.95-1.6$~$\mu$m GPI+SPHERE spectroscopy\footnote{The SPHERE and NICI photometry at $K_{\rm s}$ agree to within errors.  As a separate check on these results, we downloaded and reduced $K_{\rm s}$ data from the Keck/NIRC2 archive and derived photometry that agreed with both measurements, within errors}. Figures~\ref{fig:chi2phot} and~\ref{fig:chi2spec} display the $\chi^{2}$ distribution for our model fits to the photometry and spectroscopy, respectively. Figure~\ref{fig:sed} displays an acceptably-fitting model compared to the $YJHK_sL^{\prime}M^{\prime}$ photometry and $0.95-1.6$~$\mu$m spectroscopy. 

For fitting the photometry alone, our best fit model has a reduced chi-square $\chi^{2}_{\rm red}=$\chisqrbestphot~(d.o.f$=6-2=4$). Our 95\% cutoff for photometry fits corresponds to surface gravities of log $g=$\loggbestphot~dex, and effective temperature of $T_{\rm eff}=$\teffbestphot~$K$. Our model fits to the photometry are rather degenerate given our small number (6) of photometry points and the similarity of broadband colors for objects at $T_{\rm eff}$ = 2900--3500 $K$.

However, combining fits to the photometry and spectroscopy significantly constrains the temperature and surface gravity of HD 1160 B.  Our best-fit model for the spectrum has a reduced chi-square $\chi^{2}_{\rm red}=$\chisqrbestspec~(d.o.f$=57-2=55$)~with a surface gravity of log $g=$ 4.5~dex and effective temperature of $T_{\rm eff}=$3000~$K$.  The solutions generally agree with M16 and cluster around $T_{\rm eff}=3000--3100$~$K$ and log $g$ = 4--4.5, although we cannot rule out a 2900 $K$, log $g$ = 4.5 model nor a 3000 $K$, log $g$ = 5.0 model. We identify a suite of acceptably fitting models with a \textit{solar} metallicity, not a \textit{subsolar} metallicity.  Thus, subsolar metallicities are not required to reproduce HD 1160 B's spectrophotometry.

\subsection{Inferred Properties of HD 1160 B from Best-fitting models} 

The BT-Settl models matching HD 1160 B's photometry and spectroscopy allow us to infer a radius for the object. Compared to the default radius we used (1.2 $R_{J}$), our fits require radii larger by factors of~\normspec$\times$~and~\normphot$\times$~for the spectra and photometry respectively.  The set of parameter space fitting photometry and spectra simultaneously covers  $R=$\bestfitradius~$R_{\rm J}$ using our model fits to the HD 1160 B GPI+SPHERE spectrum, where errors in scaling the radius of our atmosphere models include the distance uncertainty error of $\approx$5\%.  HD 1160 B's luminosity, from our best-fit effective temperature and radius from the spectroscopy, is log $L/L_{\odot}=$\loglsunb~dex, in good 
agreement with log $L/L_{\odot}=-2.81\pm0.10$~dex for HD 1160 B from M16.  

\subsection{Physical Characteristics of HD 1160 C\label{sec:hd1160c}} 

We adopt an effective temperature of $T_{\rm eff}=$\teffc~for HD 1160 C using the~\sptc~SPEX/IRTF spectral type from N12. We obtain this $T_{\rm eff}$ by averaging the $T_{\rm eff}$ of an M3 and M4 $5-30$~Myr star from the relations of \cite{Pecaut13} (c.f. Table 5). Given that HD 1160 C is likely not $5-30$~Myr, the spectral type-$T_{\rm eff}$ relations in Table 5 of \cite{Pecaut13} may not apply. The spectral type-$T_{\rm eff}$ relations for $50-300$~Myr M-dwarfs (like HD 1160 C)
is expected to lie in between the relations for young stars and field stars. As an additional check, using the spectral type-$T_{\rm eff}$ relations for field stars from \cite{Pecaut13} gives an effective temperature of $T_{\rm eff}=3300\pm100$ for HD 1160 C, which is within error of our adopted $T_{\rm eff}$. 

We use our $J$ band absolute magnitude of~\absjmagc~mag and a bolometric correction of BC$_{\rm J}=$\bolocorjc~from the relations of \cite{Pecaut13} for young stars to estimate a bolometric magnitude of~\boloc~mag. We calculate the bolometric correction~BC$_{\rm J}=$\bolocorjc~by averaging the bolometric corrections of a M3 and M4 star from Table 5 of \cite{Pecaut13}. We compute a corresponding luminosity of~log~$L/L_{\odot}=$\loglsunc~dex, using a bolometric magnitude of 4.755 for the Sun\footnote{https://sites.google.com/site/mamajeksstarnotes/basic-astronomical-data-for-the-sun}, in fair agreement with log $L/L_{\odot}=-2.05\pm0.06$~dex from M16.

\section{Analysis of the HD 1160 System Age and Metallicity\label{sec:hd1160age}} 
\subsection{Motivation}
The age of the HD 1160 system is critical for interpreting the nature of the primary's two companions. Furthermore, HD 1160 provides a tight constraint on evolutionary models: the models should predict a common age given the observed effective temperatures and luminosities for three stars simultaneously. Given that HD~1160~A is on the main sequence, its age cannot be greater than the $300$~Myr main-sequence lifetime of an A$0$ star \citep[e.g.][]{Siess2000}. N12 derive a system age of $10-100$~Myr, while M16 derive a much wider age estimate of $30-300$~Myr. 

Here, we re-investigate the age of the system based on the derived $T_{\rm eff}$ and log$L/L_{\odot}$ for HD 1160 A, B and C.  For stars close to the hydrogen burning limit and brown dwarfs just below it, recent results suggest difficulties in deriving ages and masses from HR diagram positions \citep{Dupuy2009,Dupuy2010,Kraus15,Dupuy15,Dupuy16}, while ages derived from the upper main sequence are arguably preferable \citep{Soderblom2014}.  Thus, we report two possible age ranges: one considering HD 1160 A alone and one considering all three components jointly. 

\begin{figure}[ht]
\begin{center}
\includegraphics[angle=90,width=\textwidth]{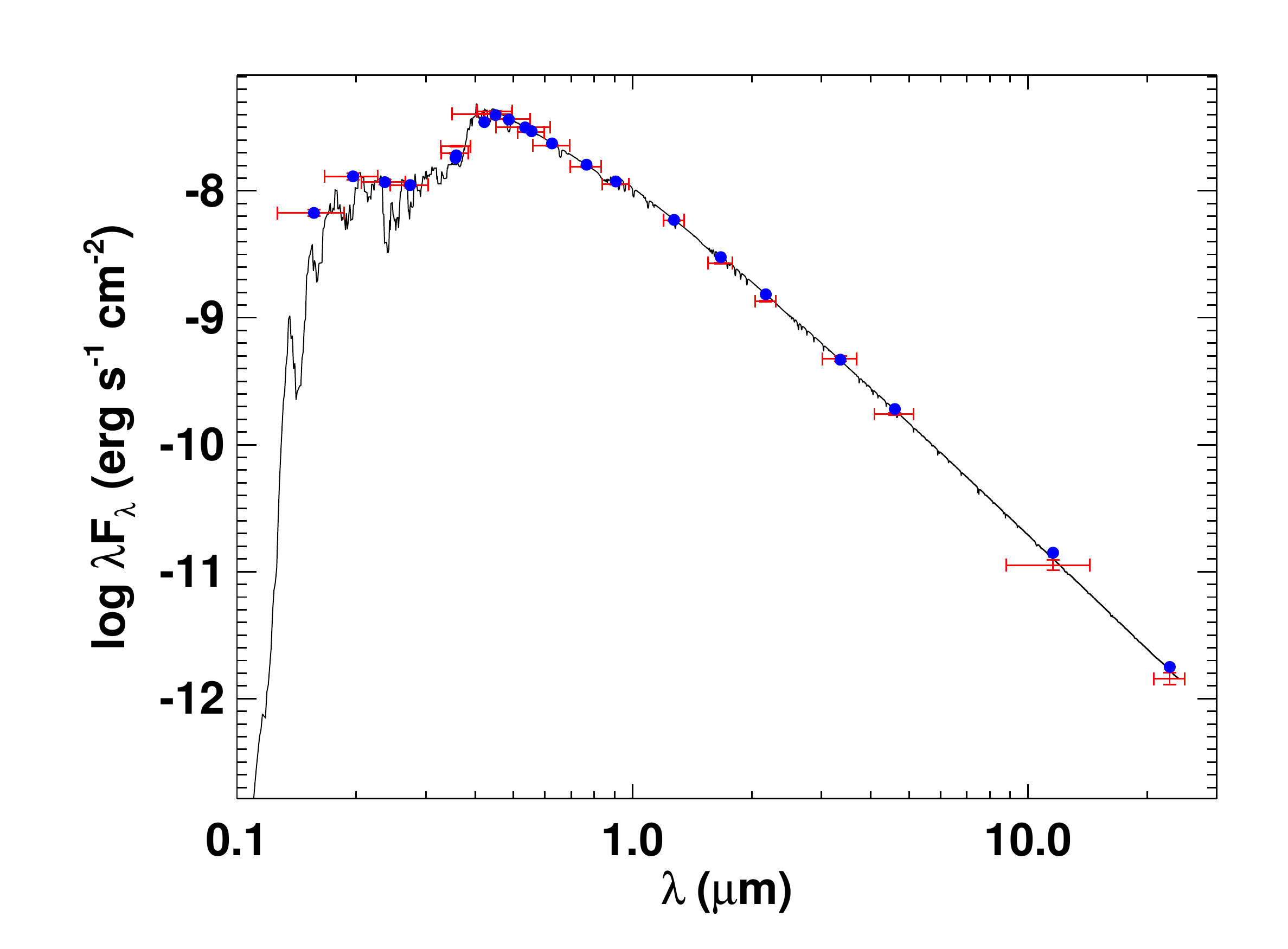}  
\end{center}
\caption{\label{fig:seda} 
SED fit to catalog photometry spanning a wavelength range of 0.16--22~$\mu$m with a NextGen atmosphere model. Red symbols are the observed fluxes with errors (horizontal `error bars' are the photometric passband widths) and blue symbols are the model fluxes. We obtain a best-fit effective temperature of $T_{\rm eff}=$\teffa~$K$ and luminosity log~$L/L_{\odot}=$\loglsuna~dex. 
}
\end{figure}

Our derived system parameters are detailed in Table 4. 
We estimate the temperature and luminosity of HD 1160 A as follows. We fit the SED of HD~1160~A using the NextGen atmosphere models of \citet{Hauschildt99}. The photometry that we fit from the literature 
spans a wavelength range of 0.16--22 $\mu$m (Figure \ref{fig:seda} and Table 5). Our best-fit model atmosphere implies an effective temperature of $T_{\rm eff}=$\teffa~K. The bolometric flux obtained by integrating the model SED provides an estimated log~$L/L_{\odot}=$\loglsuna. 

\subsection{Methodology \label{sec:modelmeth}}


To derive the range of acceptable ages, we compare the system components' luminosities and temperatures to evolutionary models. For HD 1160 A and C, we consider the Dartmouth evolutionary models \citep{Dotter08,Feiden15}, the Yonsei-Yale (Y$^{2}$) models \citep{YonseiYale03,Spada13}, and the MESA Isochrones and Stellar Tracks (MIST), which use the Modules for Experimental Astrophysics \citep[][]{Paxton11,Paxton13,Choi16,Dotter16}.  We also consider the Baraffe15 models \citep{Baraffe15} for HD 1160 B and C.  To arrive at a final age range for HD 1160 A and C, we take the median of the ranges derived from each individual evolutionary model comparison. 

All evolutionary models assume solar composition and metallicity. Our atmospheric modeling (\S\ref{sec:atmos}) suggests that HD 1160 B's spectrophotometry is consistent with a solar metallicity atmosphere. Provided that other components of HD 1160 can match solar metallicity isochrones, there is no need to posit a subsolar metallicity for the system\footnote{A substantially subsolar metallicity is unlikely for other reasons.
  Nearby young star clusters are nearly all within [Fe/H]$\approx0.1$ dex of solar metallicity \citep[see continually updated catalog from][]{Dias02}. The youngest nearby star-forming regions may be slightly sub-solar, ([Fe/H] $\approx$ -0.1) but with very little dispersion \citep{Santos2008}; slightly older moving groups like AB Dor appear slightly metal rich [Fe/H] $\approx 0.1$ \citep{Biazzo2012}. Thus, the small dispersion in metallicity amongst young groups makes [Fe/H]$= -0.3$ unlikely.}.



Given that HD 1160 A has an unknown rotational velocity v$\sin{i}$, we also consider the effect of rotation on the derived ages. Rapidly rotating stars such as Altair, Vega and substellar companion host Kappa~And A exhibit temperature changes of $\sim1000-2000$~$K$ from the stellar pole to equator. This effect is due to gravity darkening, which is directly observed via optical/IR interferometry \citep{Monnier07,Monnier12,Jones16} and modelled extensively by \cite{Espinosa11}. Given this large temperature change across the star's surface, only a rapidly rotating star's disk-integrated apparent luminosity is observable, which varies with inclination. \cite{Jones16} observed a 16\% difference in the true and apparent luminosity of Kappa And A, a difference that impacts the star's best-estimated age. HD 1160 A may be a rapidly rotating star, and thus only it's apparent luminosity and apparent effective temperature are directly observable without interferometry. Therefore, we derive a Bayesian age estimation for HD 1160 A from the \cite{BrandtHuang15} server \footnote{www.bayesianstellarparameters.info} which incorporates these uncertainties. \citet{BrandtHuang15}'s method utilizes the Geneva evolutionary models \citep{Georgy13}, the atmosphere models of \cite{Castelli04}, and orientation-dependent effective temperatures of \cite{Espinosa11}.


\newcommand{\baraffeB}{$>80$}
\newcommand{\baraffeC}{$>45$}
\newcommand{\dsepA}{$13-150$}
\newcommand{\dsepC}{$>50$}
\newcommand{\mistA}{$20-125$}
\newcommand{\mistC}{$>70$}
\newcommand{\yyA}{$20-80$}
\newcommand{\yyC}{$>75$}

\subsection{Results}
\begin{figure}[ht]
            \centering
            \includegraphics[trim={0.6cm 0.6cm 0.6cm 0.6cm},clip,angle=90,width=0.45\textwidth]{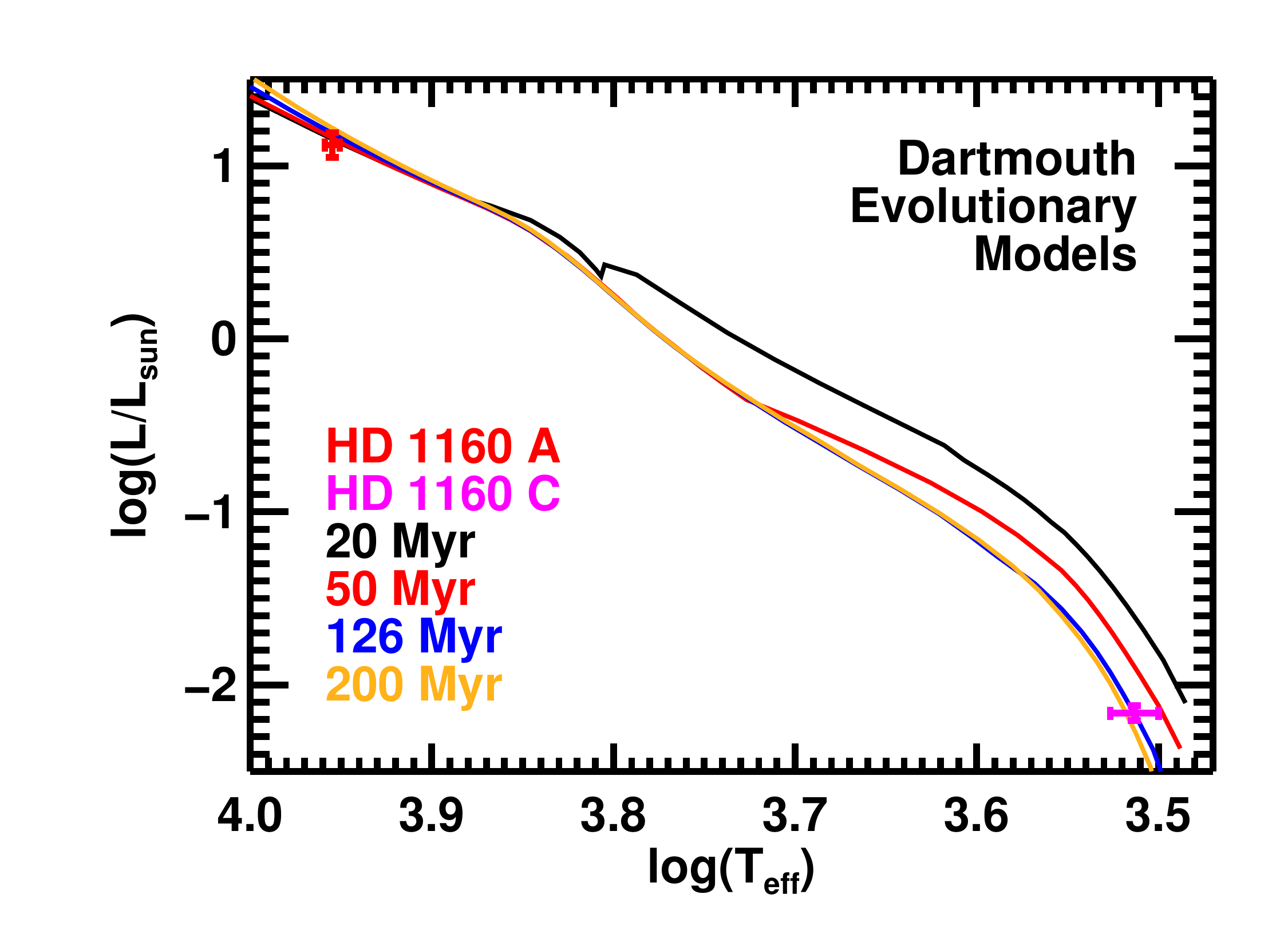}
            %
            \includegraphics[trim={0.6cm 0.6cm 0.6cm 0.6cm},clip,angle=90,width=0.45\textwidth]{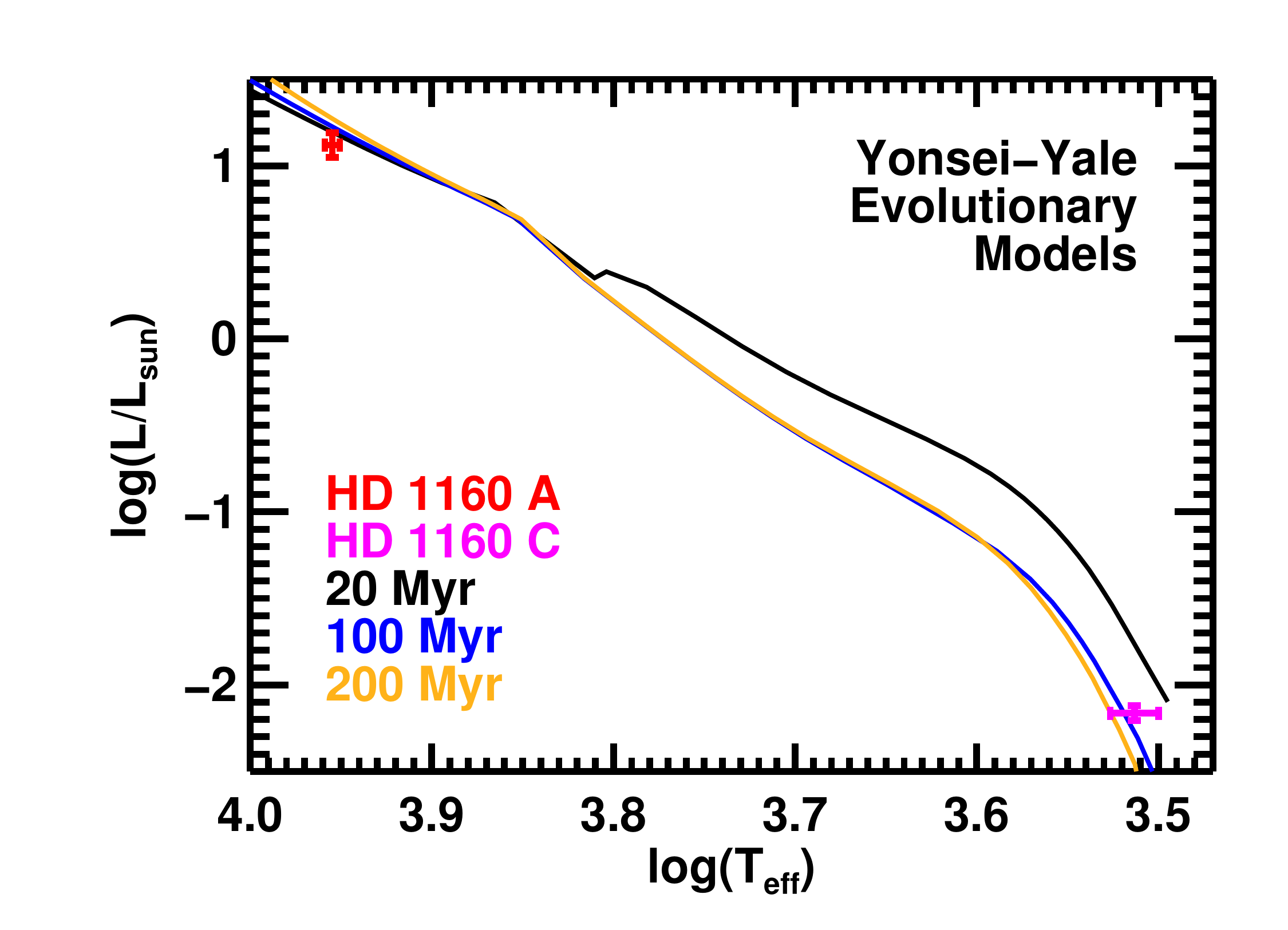}
        
        
            \includegraphics[trim={0.6cm 0.6cm 0.6cm 0.6cm},clip,angle=90,width=0.45\textwidth]{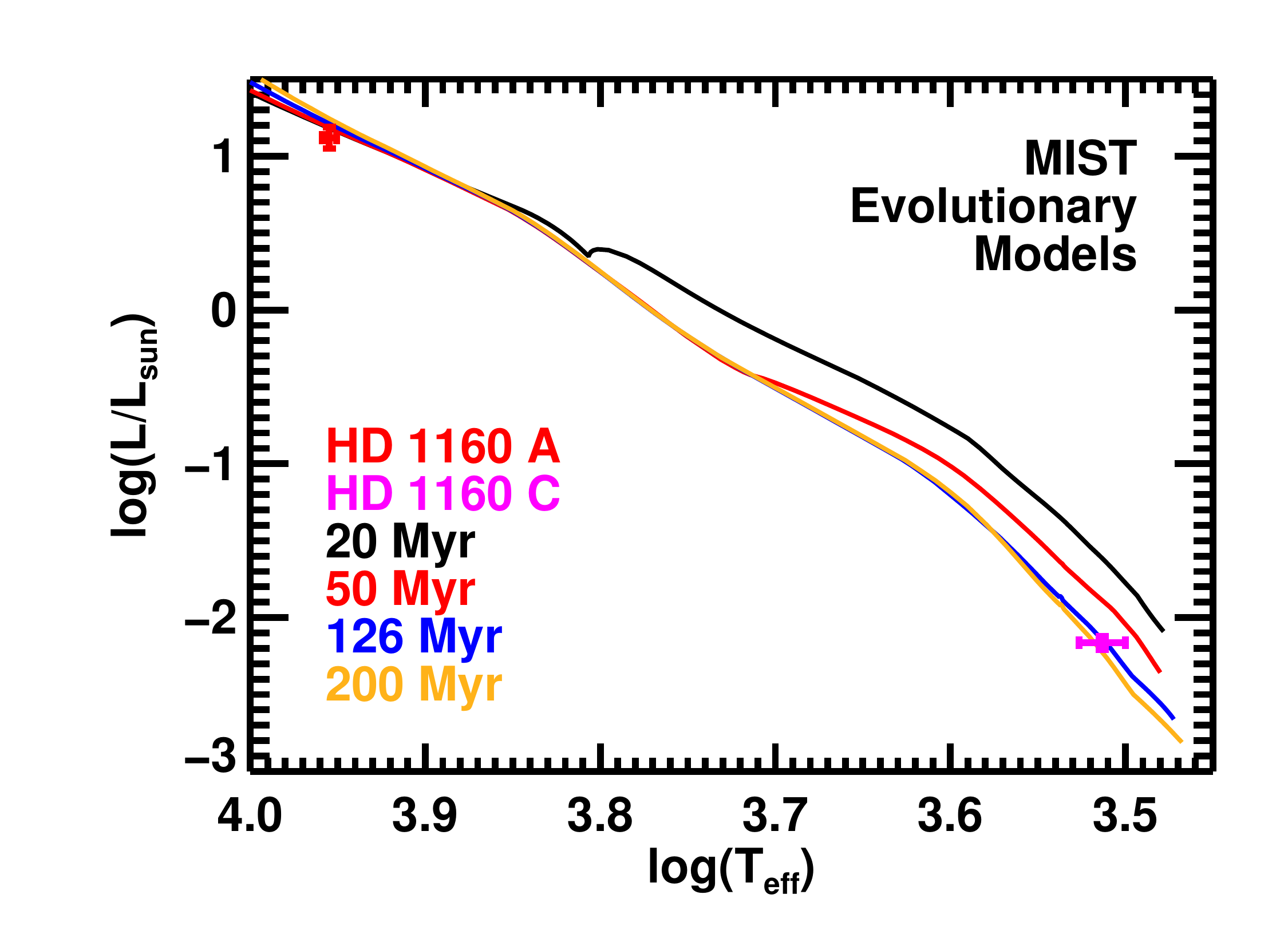}
            \includegraphics[trim={0.6cm 0.6cm 0.6cm 0.6cm},clip,angle=90,width=0.45\textwidth]{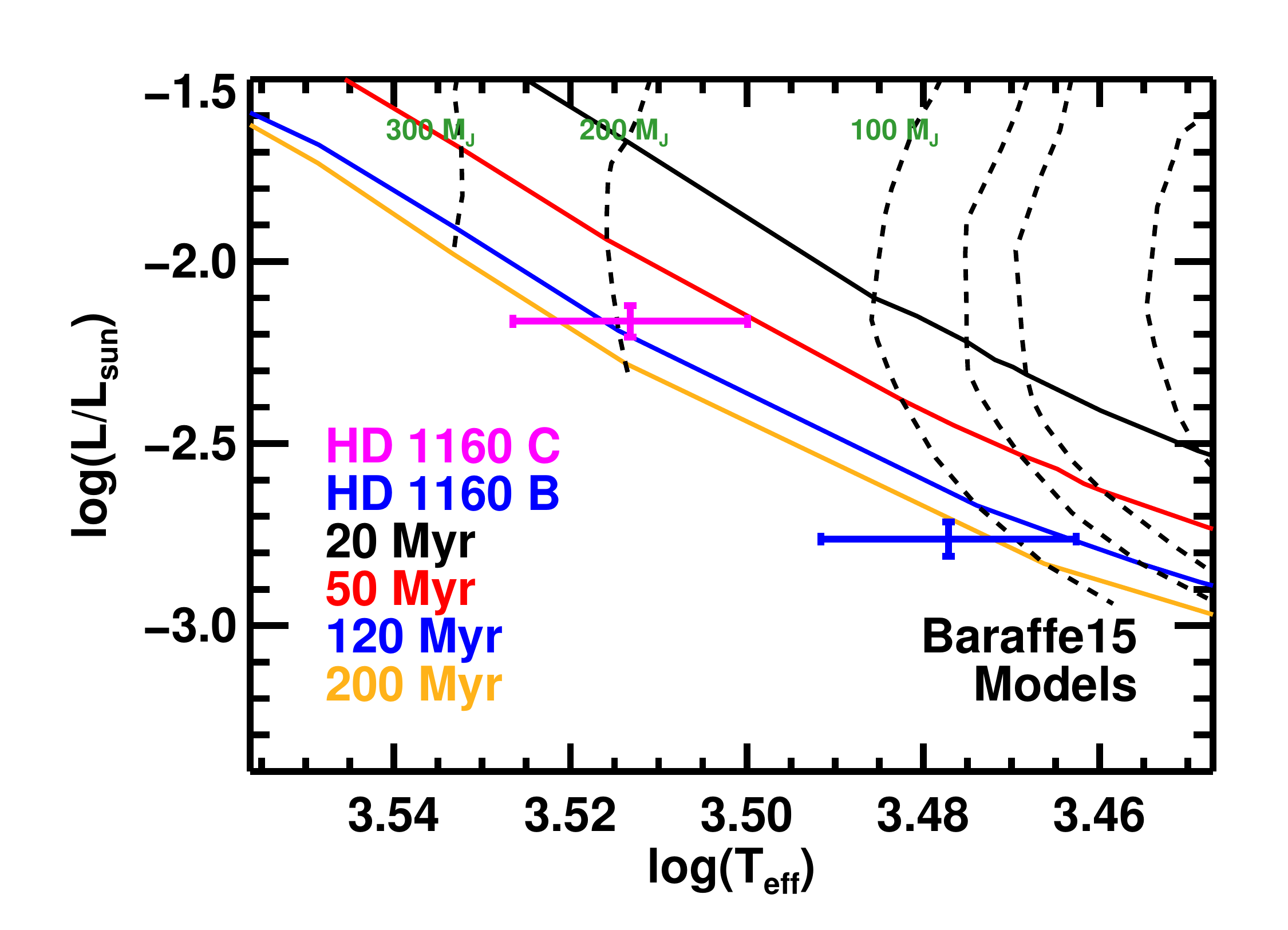}
        \caption[ CMD ]
        {\label{fig:models} A comparison of the effective temperature and luminosities of HD~1160~A (red point, derived from SED fitting in Figure~\ref{fig:seda}), and HD 1160 C (light blue point, see~\S\ref{sec:hd1160c}) to stellar models. The Dartmouth, Yonsei-Yale and MIST evolutionary models do not extend into the sub-stellar ($<0.1$~$M_{\odot}$) regime, precluding a reliable age estimate for HD 1160 B, while the Baraffe15 models do not extend to the mass of HD 1160 A.  
        }
        
    \end{figure}

\begin{figure}[ht]
\begin{center}
\includegraphics[scale=0.75]{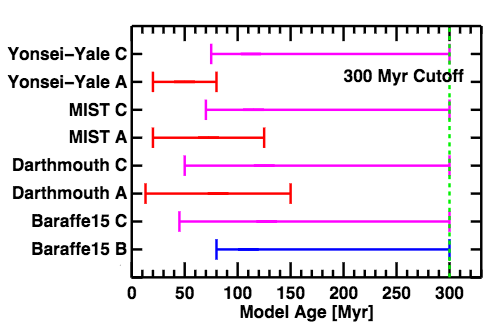}  
\end{center}
\caption{\label{fig:ageplot} 
Ages derived from the Baraffe15, DSEP, MIST, and Yonsei-Yale isochrones (see Figure~\ref{fig:models}) for HD 1160 A, B and C. The model ages are derived from the observed $T_{\rm eff}$ and log~$L/L_{\odot}$ for each star. The bright green dotted line denotes the 300~Myr cut off age for the system, given that HD 1160 A is a main-sequence star \citep{Siess2000}. 
}
\end{figure}

Our comparisons of HD 1160 to evolutionary models are shown in Figures~\ref{fig:models} and~\ref{fig:ageplot}.  Evolutionary models suggest a young age for HD 1160 A: \dsepA~Myr, \yyA~Myr, and \mistA~Myr for the Dartmouth, Yale-Yonsei, and MIST models, respectively, where we quote age ranges consistent with 1-$\sigma$ uncertainties in temperature and luminosity.  The Bayesian techniques of \cite{BrandtHuang15} also point towards a young age.   Bayesian methods are consistent with this estimate, with a `best-estimated' age (mean of the age$\times$probability) of $\sim 58$ Myr, and reject ages $\gtrsim$ 200 Myr at 90\% confidence.     

Model comparisons to HD 1160's lower-mass components generally favor older ages.  The Dartmouth, Yale-Yonsei, MIST and Baraffe models favor ages of  \dsepC~Myr, \yyC~Myr, \mistC~Myr, and \baraffeC~Myr for HD 1160 C.   The Baraffe models imply an age of \baraffeB~Myr for HD 1160 B.  In most cases, the lower limit on the components' luminosity implies stellar objects on/near the main sequence, precluding us from assigning age upper limits.

Therefore, considering HD 1160 A alone, we derive an age range of $\approx$ 20--125 Myr for the system.
Considering HD 1160 ABC jointly implies a system age of 80--125~Myr.  That much of the model age ranges for HD 1160 A vs. HD 1160 BC do not overlap at the 1-$\sigma$ level (Fig. \ref{fig:ageplot}) may suggest discrepancies between models and the true physical properties of young M-dwarfs \citep{Feiden15,Kraus15}.


\section{Masses of HD 1160 B and C}

Given an age of $80-125$~Myr from analyzing HD 1160 A, B and C jointly, and luminosities of log~$L/L_{\odot}= -2.76\pm0.05$~and log~$L/L_{\odot}=$\loglsunc, we derive mass ranges of~\massb~$M_{\rm J}$ and~\massc~$M_{\rm J}$ for HD 1160 B and C respectively, using the \cite{Baraffe15} models.  Considering instead the age range derived from HD 1160 A alone, we estimate masses of 35--90 $M_{\rm J}$ for B and 110--200 $M_{\rm J}$ for HD 1160 C. 

Mass estimates derived from our atmospheric modeling are broadly consistent with those derived from luminosity evolution.  
The best-fit radius and surface gravity for HD 1160 B in turn implies a mass of $\sim$ 29$^{+63.8}_{-20.8}$ $M_{\rm J}$.  Surface gravity is the key factor, as models with log $g$ = 4.0 imply a planet-mass companion (8--9 $M_{\rm J}$), those with log $g$ = 4.5 imply a $\sim$ 30 $M_{\rm J}$ brown dwarf and those with log $g$ = 5.0 imply an object slightly above the hydrogen burning limit of $70-80$~$M_{\rm J}$\citep{Dieterich14} \footnote{As point of comparison, the surface gravity predicted for HD 1160 B by the Baraffe models for a 30 $Myr$-old, 35 $M_{\rm J}$ object is log~$g$ $\sim$ 4.4 and fpr a 125 $Myr$-old, 90 $M_{\rm J}$ object is log~$g$ $\sim$ 4.9.  While a subset of these surface gravities are consistent with results from our atmospheric modeling, the predicted temperatures are significantly lower (2600--2850 $K$), discrepancies that may point to additional challenges in predicting the evolution of young substellar objects and the lowest-mass stars \citep[see also][]{Kraus15}}. 

Thus, the interpretation of HD 1160 B changes depending on which system component(s) we use to estimate an age.  Using only HD 1160 A to calibrate the system's age, HD 1160 B's allowable mass range mostly covers the substellar regime.  But ages derived from both HD 1160 A, B and C preclude the youngest ages (20--80 Myr), and suggest that HD 1160 B straddles the hydrogen burning limit.  

Our mass ranges for HD 1160 B agree with those of M16, although our mass upper limit is smaller due to our somewhat lower due to upper age limit of $\sim125$~Myr for the HD 1160 system. However, our mass ranges for HD 1160 B are wider than the mass range of $24-45$~$M_{\rm J}$ from N12 using $J$-band photometry, and generally skew towards larger values. 


\section{Conclusions} 

We present new and precise SCExAO near-infrared photometry, reprocessed thermal infrared photometry, and $Y$ band GPI integral field spectroscopy of the HD 1160 system, which hosts two low-mass companions discovered by N12.  Combining our data with recently-published near-IR spectroscopy from SPHERE, we constrain the infrared colors of HD 1160 BC and constrain the atmospheric properties of HD 1160 B.  After revisiting the age of the HD 1160 system using multiple diagnostics, we determine the likely intrinsic properties of HD 1160 B and C.  

Our study yields the following results:

\begin{itemize}
    \item Our SCExAO photometry refines estimates for HD 1160 B's brightness in major infrared passbands.  Our photometry in general agrees with the SPHERE $1.0-1.6$~$\mu$m spectrum of M16, although it is systematically fainter by $15-25$\% for HD 1160 B. Like M16, our SCExAO photometry for HD 1160 B is in strong disagreement with the $J$ and $H$ photometry of~\cite{Nielsen12}~by 1.14 and 0.44 mag respectively; our re-derived $M^\prime$ photometry likewise shows differences. \item These differences, in turn, lead to a reinterpretation of HD 1160 B and C's infrared colors.  Specifically, HD 1160 BC are not discrepant or redder than the field sequence but completely consistent with the colors expected for field dwarfs with early-to-mid M spectral types.
    \item We revise the spectral type of HD 1160 B to M5.5$^{+1.0}_{-0.5}$, earlier than proposed by N12 based purely on near-infrared colors but generally consistent with the results from M16 based on longer-wavelength spectra.
    \item Our atmospheric modeling finds a best-fit temperature for HD 1160 B of 3000-3100 $K$ and a surface gravity of log $g$ = 4--4.5, although we cannot rule out slightly cooler temperatures or slightly higher surface gravities over some portion of parameter space.  
    \item In contrast with the results from M16, there is no evidence from atmospheric modeling for HD 1160 B having a subsolar metallicity, as the best-fitting (solar metallicity) models are consistent with the data, having a reduced $\chi^{2}$ $\sim$ 1.
    \item We find different answers for the system's age depending on which components are analyzed.  Considering HD 1160 A, B and C jointly, we estimate an age of 80--125 Myr, implying masses of 70--90 $M_{\rm J}$ for HD 1160 B, making it an object straddling the hydrogen burning limit.  Considering HD 1160 A alone, we estimate an age of 20--125 Myr, a wider range than that listed in N12 and younger than that estimated in M16.  Given this age, we derive a mass of 35--90 $M_{\rm J}$ for HD 1160 B.  Estimated masses for HD 1160 B from atmospheric modeling are consistent with these ranges but allow for slightly lower masses.  Our analysis is then consistent with HD 1160 B being a relatively intermediate-to-high mass brown dwarf and HD 1160 C being a low-mass star. 
    
\end{itemize}

HD 1160 is an important triple system, with one high-mass star, one very low-mass star, and one brown dwarf/extremely low mass star.  As such, it presents an excellent laboratory for studying (sub)stellar evolution and testing our understanding of luminosity evolution.  Our results may hint at challenges ahead, as age estimates derived from HD 1160 A, although formally consistent with those from HD 1160 BC, tend to be younger than those derived from the system's lower-mass companions.  This trend is an inversion of that seen for ages derived for the youngest clusters/moving groups: e.g. ages derived for Upper Scorpius intermediate-mass stars are  older than those derived for its lower-mass stars \citep[e.g.][]{Pecaut12, Preibisch02}.   Interferometric measurements of HD 1160 A's angular diameter, combined with a GAIA parallax more precise than Hipparcos, would clarify the primary's true luminosity and temperature, allowing us to place more firm constraints on its age based on its HR diagram position.

HD 1160 B is one of many young substellar objects that may provide key insights into young substellar object formation and atmospheres.  Its mass ratio ($\sim$ 0.015--0.040 or 0.027--0.040) perhaps indicates formation by protostellar disk fragmentation/disk instability \citep[e.g.][]{Boss2011}, or is consistent with the lowest-mass objects formed like binary stars \citep{Reggiani2011}.  Indeed, HD 1160 B's mass range and the semi-major axis is compatible with a disk-instability scenario of formation, provided the disk is $\sim50\%$ the star's mass \citep[Figure C.1 in][]{Bonnefoy14}. 

As shown by works like \citet{Currie13,Currie14a} and most recently (and exhaustively) by \citet{Faherty16}, low-gravity L- and T-type brown dwarfs and directly imaged planets across a wide range of spectral types and temperatures appear to depart from the field dwarf sequence.   HD 1160 B, an earlier spectral type object, does not yet show differences.  Identifying where/why young and low-gravity objects begin to depart from the field provides some insight into and empirical constraint on the evolution of substellar atmospheres.  

While the masses inferred for HD 1160 B from atmospheric modeling appear consistent with those implied by luminosity evolution models, our results hint that follow-up spectroscopy could better constrain the object's surface gravity and (by inference) mass.  The implied mass of HD 1160 B varies by a factor of ten from the lowest to highest surface gravity from our suite of acceptably fitting models.  As described in \citet{Allers13}, higher resolution spectroscopy focused on line transitions such as Na I and K I could better clarify how HD 1160 B's surface gravity compares to that of other brown dwarfs with a range of ages.  Facilities such as Keck/OSIRIS and VLT/SINFONI could potentially provide these data.

\begin{table}[htbp]
 \begin{center}
Table 1\\
Observation log of HD 1160\\
 \begin{tabular}{llllllll}
\hline
\hline
Telescope & Instrument & UT date & Band & Wavelength & $t_{\rm int}$ & $N_{\rm exp}$ & Mode \\
          &            &         &      & ($\mu$m)     & (s)               & \\
\hline 
Subaru       & SCExAO/HiCIAO & 10-31-2015 & $Y$ & \scexaoywave & 1.5 & 11 & Photometry \\
Subaru       & SCExAO/HiCIAO & " & $Y$+ND & \scexaoywave & 1.5 & 6 & Photometry \\
Subaru       & SCExAO/HiCIAO & " & $J$ & \scexaojwave & 1.5 & 5 & Photometry \\ 
Subaru       & SCExAO/HiCIAO & " & $J$+ND& \scexaojwave & 1.5 & 5 & Photometry \\ 
Subaru       & SCExAO/HiCIAO & " & $H$ & \scexaohwave & 1.5 & 5 & Photometry \\
Subaru       & SCExAO/HiCIAO & " & $H$+ND & \scexaohwave & 1.5 & 5 & Photometry \\ 
Gemini-South & GPI/IFS       & 11-17-2013 & $Y$ & \gpiwave & 88.7 & 9 & IFS \\
\hline
 \end{tabular}
 \end{center}
 \caption*{Observation log of HD 1160.  ND stands for ``neutral density filter": these are flux calibration observations obtained for HD 1160 with the primary unsaturated and in the linear count regime. The neutral density filters are~\ndy,~\ndj~and~\ndh~for $Y$,$J$ and $H$, respectively.}
 \label{table:obslog}
 \end{table}

\clearpage
\begin{table}[htbp]
 \begin{center}
Table 2\\
\setlength{\tabcolsep}{0.06in}
Near-to-Mid IR Photometry of HD 1160 B and C\\
 \begin{tabular}{lllllll}
\hline
\hline
Object    & Band   & SCExAO/HiCIAO & GEMINI/GPI & VLT/SPHERE & Gemini/NICI & Keck/NIRC2  \\
          & (MKO)  & (mag)         & (mag)      & (mag)      & (mag)  & (mag)\\
\hline 
HD 1160 B &  $Y$    &\scexaoyb$^1$      & \gpiyb$^{1,4}$     &      & &\\
          &  $J$    &\scexaojb$^1$      &               & \vltjb$^2$     & \nicijb$^3$ &\\
          &  $H$    &\scexaohb$^1$      &               & \vlthb$^2$     & \nicihb$^3$ &\\ 
          &  $K_s$    &                   &              &  \vltkb$^2$             & \nicikb$^3$& \\
          & $M^\prime$ &                &               &                 &    &\keckmb$^1$\\      
\hline 
HD 1160 C &  $Y$    & \scexaoyc$^1$     &                &           &\\
          &  $J$    & \scexaojc$^1$     &                &    & \nicijc$^3$ \\
          &  $H$    & \scexaohc$^1$     &               &            & \nicihc$^3$ \\ 
          &  $K_s$    &                   &               &            & \nicikc$^3$ \\
\hline
 \end{tabular}
 \end{center}
  \caption*{Photometry. \\
$^1$This work. \\
 $^2$ The SPHERE $J$ and $H$ photometry is estimated from the IFS spectrum over the standard MKO $J$ and $H$ filter bandpasses.  The $H$ band portion of the SPHERE spectrum is incomplete, covering only the blue half of the standard MKO $H$ band wavelength range.  The $K_{\rm s}$ band photometry is adopted from \cite{Maire15}.  Note that we also adopt their $L^\prime$ measurement of m($L^\prime$) = 13.60 $\pm$ 0.10 for HD 1160 B. \\
 $^3$\cite{Nielsen12} \\
 $^4$Note that the GPI $Y$ band filter extends to slightly longer wavelengths than HiCIAO.
    }
 \label{table:phot}
 \end{table}

\begin{table}[htbp]
 \begin{center}
Table 3\\
Model Fitting Results\\
 \begin{tabular}{lll}
\hline
\hline
                                                 & Photometry                    & Spectroscopy \\
Wavelength Range                                 & $YJHK_sL^{\prime}M^{\prime}$  & 0.95-1.6~$\mu$m \\ 
$T_{\rm eff}$ K (95\% Confidence)                & \teffbestphot                 &  \teffb $K$ \\
                                                 &                   & 2900 $K$ (for log $g$ = 4.5)\\
log $g$ dex (95\% Confidence)                     & \loggbestphot                 & \loggb    \\
                                                &                               & 5.0 (for 3000 $K$)\\
Models fitting both       & $T_{\rm eff}=$\teffb, log $g=$\loggb\\
Photometry and Spectra: & $T_{\rm eff}=$2900~$K$, log $g=4.5$; \\
&  $T_{\rm eff}=$3000~$K$, log $g=5.0$ \\
\hline 
 \end{tabular}
 \end{center}
 \label{table:sed}
 \end{table}

\newcommand{\ra}{$00:15:57.3025$}
\newcommand{\dec}{$+04:15:04.018$}
\newcommand{\propra}{$21.15\pm0.62$}
\newcommand{\propdec}{$-14.20\pm0.24$}
\newcommand{\age}{$20-125$} 

\newcommand{\distancemod}{$5.07\pm0.10$}
\newcommand{\absmagya}{$1.89\pm0.10$}
\newcommand{\absmagja}{$1.91\pm0.10$}
\newcommand{\absmagha}{$1.94\pm0.10$}
\newcommand{\absmagka}{$1.97\pm0.10$}

\newcommand{\appmagya}{$6.96\pm0.02$} 
\newcommand{\appmagja}{$6.98\pm0.02$} 
\newcommand{\appmagha}{$7.01\pm0.02$} 
\newcommand{\appmagksa}{$7.04\pm0.03$} 
\newcommand{\appmaglpa}{$7.055\pm0.014$} 
\newcommand{\appmagmsa}{$7.04\pm0.02$} 

\newcommand{\absmagkc}{$7.15\pm0.12$}
\newcommand{\deltamagkc}{$5.14\pm0.06$}

\newcommand{\absmagk}{$8.88\pm0.12$}
\newcommand{\absmagknici}{$9.09\pm0.12$}
\newcommand{\deltamagk}{$6.91\pm0.10$}

\begin{table}[htbp]
 \begin{center}
 \setcounter{table}{3}
Table 4\\
Properties of the HD 1160 System\\
 \begin{tabular}{ccccc}
\hline
Property          & HD~1160~A   &   HD 1160 B        & HD 1160 C        & Unit \\
R.A. (ep J2000)    & \multicolumn{3}{c}{\ra$^{4}$}                             & \\
Dec. (ep J2000)    & \multicolumn{3}{c}{\dec$^{4}$}                             & \\
Distance          & \multicolumn{3}{c}{\distance$^{3}$}                 & pc \\
$\mu_{\alpha}$    & \multicolumn{3}{c}{\propra$^{4}$}                 & mas yr$^{-1}$\\
$\mu_{\delta}$ & \multicolumn{3}{c}{\propdec$^{4}$}                 & mas yr$^{-1}$\\
Age, HD 1160 A    & \multicolumn{3}{c}{$20-125$}                            & Myr \\
Age, HD 1160 A, B and C & \multicolumn{3}{c}{$80-125$}                            & Myr \\

\hline
$M_{Y}$(MKO)  &\absmagya$^6$& \absmagy           & \absymagc        & mag \\
$M_{J}$       & \absmagja   & \absmagj           & \absjmagc        & mag \\
$M_{H}$       & \absmagha   & \absmagh           & \abshmagc        & mag \\
$M_{Ks}$      & \absmagka   & \absmagk$^5$       & \absmagkc$^1$    & mag \\
$\Delta Y$ (MKO)   &             & \deltamagy         & \deltamagyc      & mag\\
$\Delta J$         &             & \deltamagj         & \deltamagjc      & mag\\
$\Delta H$         &             & \deltamagh         & \deltamaghc      & mag \\
$\Delta K$         &             & \deltamagk$^5$     & \deltamagkc$^1$  & mag \\
SpT               & \spta$^{1}$ & \sptb              & \sptc$^{1}$      & \\
$T_{\rm eff}$     & \teffa      & \teffb             & \teffc$^{2}$     & K \\
Log$L/L_{\odot}$  & \loglsuna   & \loglsunb          & \loglsunc        & dex \\
Log$(g)$          & \logga      & \loggb             & \loggc           & dex \\
Mass              &  2.2$^{1}$    &                    &                  & $M_{\odot}$      \\ 
                  &             & 35--90$^{7}$             & 110--200$^{7}$           & $M_{\rm J}$ \\
                  &             & \massb$^{8}$             & \massc$^{8}$           & $M_{\rm J}$ \\

\hline
 \end{tabular}
 \end{center}
 \caption*{ $^{1}$\cite{Nielsen12}. \\$^{2}$\cite{Maire15}. \\$^{3}$ \cite{vanLeeuwen07}. \\$^{4}$\cite{vanLeeuwen07}. \\
 $^5$ Calculated by \cite{Maire15} at the K$_s$ passband using the SPHERE spectrum and IRDIS K$_1$ and K$_2$ photometry.\\  $^6$ Calculated from our SED fit to HD~1160~A photometry (see \S\ref{sec:hd1160age}). \\$^{7}$ mass ranges from \cite{Baraffe15} models using an age derived for HD 1160 A alone (see \S\ref{sec:hd1160age}). \\$^{8}$ mass ranges from \cite{Baraffe15} using an age derived for HD 1160 A, B and C (see \S\ref{sec:hd1160age}).
 \label{table:absmag}}
 \end{table}
 
 \clearpage

 \begin{table}[htbp]
 \begin{center}
Table 5\\
Observed Fluxes of the Spectral Energy Distribution of HD~1160~A\\
 \begin{tabular}{lll}
\hline
\hline
Wavelength   & Width       & Flux  \\
$[\mu$m$]$   & $[\mu$m$]$  & [ergs/s/cm$^{-2}$ $\times10^{-10}$] \\
\hline
$0.274$ & $0.030$ & $111.000\pm2.740$\\
$0.236$ & $0.030$ & $117.000\pm6.150$\\
$0.197$ & $0.030$ & $130.000\pm8.250$\\
$0.157$ & $0.030$ & $67.000\pm4.230$\\
$0.422$ & $0.145$ & $403.361\pm3.715$\\
$0.535$ & $0.167$ & $317.359\pm2.923$\\
$0.358$ & $0.062$ & $224.376\pm2.067$\\
$0.450$ & $0.091$ & $423.056\pm3.896$\\
$0.556$ & $0.086$ & $289.911\pm2.670$\\
$0.357$ & $0.056$ & $198.323\pm1.827$\\
$0.487$ & $0.129$ & $367.277\pm3.383$\\
$0.626$ & $0.134$ & $227.942\pm2.099$\\
$0.765$ & $0.137$ & $155.036\pm1.428$\\
$0.907$ & $0.140$ & $112.999\pm1.041$\\
$1.273$ & $0.152$ & $58.606\pm0.540$\\
$1.671$ & $0.237$ & $26.964\pm0.248$\\
$2.171$ & $0.255$ & $13.514\pm0.124$\\
$3.353$ & $0.663$ & $4.753\pm0.232$\\
$4.603$ & $1.042$ & $1.745\pm0.031$\\
$11.561$ & $5.507$ & $0.112\pm0.010$\\
$22.800$ & $4.101$ & $0.014\pm0.002$\\
\hline
 \end{tabular}
 \end{center}
 \tablecaption{Fluxes from the literature used in the SED fitting of HD 1160 A, from \cite{Thompson78},  \cite{Cutri03}, \cite{Ammons06}, \cite{Pickles10}, and \cite{Cutri12}. Column 3 are the red data points in Figure~\ref{fig:seda}.\label{table:seda}
 }
 \label{table:fluxes}
 \end{table}

\begin{table}[htbp]
 \begin{center}
Table \ref{table:age}\\
Age estimations of the HD 1160 System\\
 \begin{tabular}{ll}
\hline
\hline
Evidence  & Age Constraint  \\
\hline 
HD~1160~A is a main-sequence early A star             & $\lesssim300$~Myr \\
Darthmouth Models                                  & \dsepA~Myr for HD~1160~A; \dsepC~Myr for HD 1160 C  \\ 
Yonsei-Yale Models                                 & \yyA~Myr for A; \yyC~Myr for C  \\
Baraffe Models                                     & \baraffeB~Myr for B; \baraffeC~Myr for C\\
MIST Models                                        & \mistA~Myr for A;  and \mistC~Myr for C \\
Bayesian analysis of Geneva Models                 & $\sim$ 58 Myr (best-estimated age)\\
& $\lesssim200$ Myr ($90\%$ confidence) for HD~1160~A \\
Adopted Age HD 1160 A only                                       &  20--125 Myr\\
Adopted Age HD 1160 A, B and C                                      &  80--125 Myr\\
 \end{tabular}
 \end{center}
 \tablecaption{Ages of HD~1160 system as shown in Figure~\ref{fig:ageplot} and detailed in \S\ref{sec:hd1160age}. }
 \label{table:age}
 \end{table}

\begin{figure}[ht]
\begin{center}
\includegraphics[angle=90,width=\textwidth]{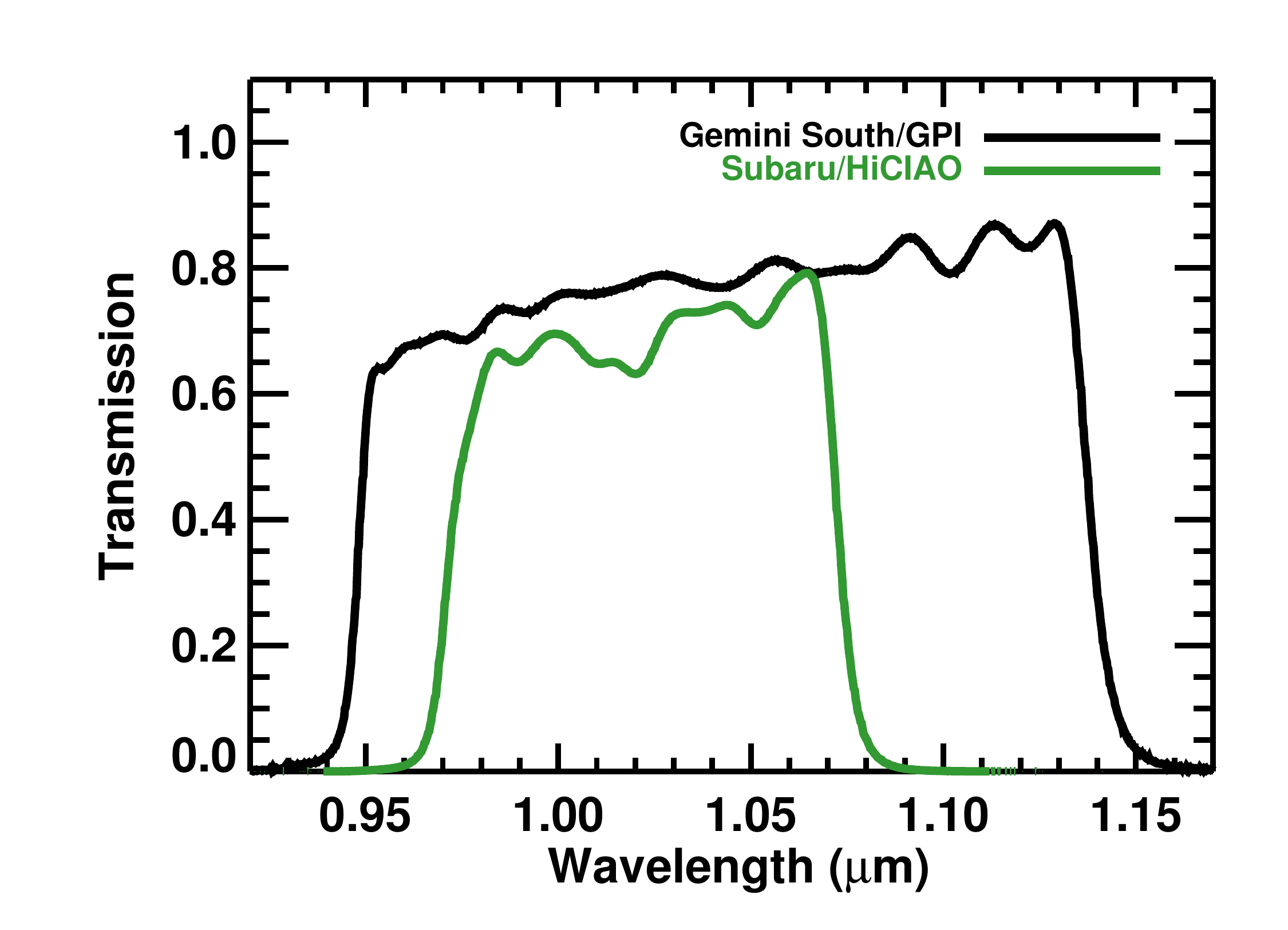}  
\end{center}
\caption{\label{fig:filt} A comparison of the GPI $Y$ band filter and 
the HiCIAO $Y$ band filter. The GPI filter is slightly wider on both the blue and red end in comparison to the HiCIAO filter. This leads to slightly brighter photometry for GPI $Y$ band as compared to our SCExAO/HiCIAO photometry for HD 1160 B.  
}
\end{figure}

\acknowledgments
We thank Anna-Lise Marie for detailed, helpful discussions about and access to SPHERE HD 1160 B spectra.  We thank Eric Mamajek for discussions about the metallicity of nearby, young stars. We thank Davy Kirkpatrick for providing NIR spectra of M-dwarf spectral standards. We thank Federico Spada for discussions and tests of the Yonsei-Yale models. EVG would like to acknowledge the gracious support of his Lowell Pre-doctoral Fellowship by the BF foundation and the excellent support of Fisk-Vanderbilt Bridge Program. This research has made use of the SIMBAD database, operated at CDS, Strasbourg, France. This research has made use of the Keck Observatory Archive (KOA), which is operated by the W. M. Keck Observatory and the NASA Exoplanet Science Institute (NExScI), under contract with the National Aeronautics and Space Administration.  The authors acknowledge support from the JSPS (Grant-in-Aid for Research \#$23103002$, \#$23340051$, and \#$26220704$). This work was partially supported by the Grant-in-Aid for JSPS fellows (Grant 
Number 25-8826). M. Janson acknowledges support of the U.S. National Science Foundation, under Award No. 1009203. This work performed under the auspices of the U.S. Department of Energy by Lawrence Livermore National Laboratory under Contract DE-AC52-07NA27344 with document release number LLNL-JRNL-701012-DRAFT. We wish to acknowledge the pivotal cultural role and reverence that the summit of Maunakea has always had within the indigenous Hawaiian community.  We are most fortunate to have the privilege to conduct scientific observations from this mountain.

\clearpage

\clearpage

\end{document}